\documentclass{IEEEtran}
\usepackage{amsmath,amsfonts}
\usepackage{algorithmic}
\usepackage{algorithm}
\usepackage{array}
\usepackage[caption=false,font=normalsize,labelfont=sf,textfont=sf]{subfig}
\usepackage{textcomp}
\usepackage{stfloats}
\usepackage{url}
\usepackage{verbatim}
\usepackage{graphicx}
\usepackage{cite}
\usepackage{booktabs}
\usepackage{multirow}
\begin{document}

\title{Real-Time Patient Monitoring with Heterogeneous Systems Using DDS-Based Communication}

\author{Muhammad Dikko Gambo, MD Sakibul Islam, Basem Almadani,
Farouq Aliyu,~\IEEEmembership{Senior Member,~IEEE,} Abdullahi Sani Shuaibu,~\IEEEmembership{Student Member,~IEEE,}
\thanks{Muhammad Dikko Gambo, MD Sakibul Islam, and Abdullahi Sani Shuaibu are with the Department of Computer Engineering, King Fahd University of Petroleum \& Minerals, Dhahran, Saudi Arabia.
Basem Almadani, and Farouq Aliyu are with the Department of Computer Engineering, and the Applied Research Center (ARC) for Non-Profit and Social Development, Dhahran, Saudi Arabia.}
\thanks{Manuscript received XX XX}}


\IEEEpubid{0000--0000/00\$00.00~\copyright~2021 IEEE}

\maketitle

\begin{abstract}
Real-time patient monitoring requires communication systems that maintain low latency and high reliability while scaling across heterogeneous hospital deployments. This paper presents a middleware-based monitoring system that uses the Data Distribution Service (DDS) to coordinate data exchange among distributed medical components. The system architecture consists of modular DDS domain participants deployed across patient rooms and ward-level applications, connected through a layered data bus structure. Quality of Service (QoS) policies, including Reliable and Best Effort, are configured and evaluated to examine trade-offs between delivery guarantees and communication overhead. A prototype implementation is developed to emulate clinical monitoring workflows, and experiments are conducted in comparison with socket-based messaging. The evaluation indicates that DDS with Reliable QoS avoids packet loss in the tested scenarios and provides more dependable delivery than sockets under network load. These results support the use of DDS as a practical middleware option for real-time clinical communication where consistent data delivery is required.
\end{abstract}


\section{Introduction}
\IEEEPARstart{T}{he} demand for healthcare services continues to increase with population growth, while the United Nations Sustainable Development Goals emphasize the need for accessible, high-quality, and cost-effective medical care. Advances in sensing, networking, and computing have enabled the development of remote patient monitoring systems that extend care beyond traditional clinical settings. Despite this progress, several challenges remain, including achieving dependable data delivery compared to periodic nurse- or technician-based monitoring, reducing system integration time, and managing the complexity of heterogeneous medical equipment \cite{kakria2015real}. Many deployed healthcare systems rely on conventional web-based technologies. For example, Almadani et al.~\cite{2Almadani2016} proposed a service-oriented architecture (SOA)-based e-healthcare system using open standards such as XML and SOAP to improve interoperability across platforms and programming languages. While such approaches support modular integration, they typically provide limited support for real-time communication. Other Internet Protocol-based solutions designed for mission-critical domains, such as Command, Control, Communications, Computers, and Intelligence (C4I) systems \cite{Baek2021}, offer secure communication but are often complex to deploy and maintain in healthcare environments.

\begin{figure}
    \centering
    \includegraphics[width=0.475\textwidth]{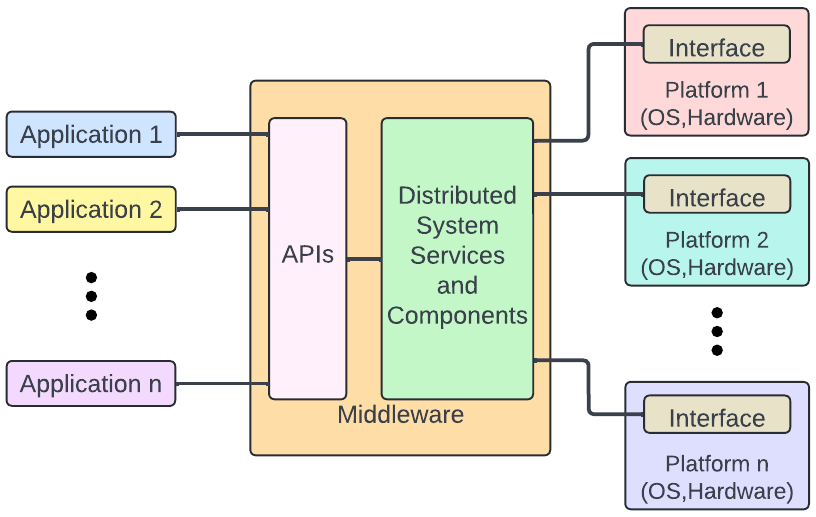}
    \caption{A basic middleware architecture}
    \label{fig:Middleware}
\end{figure}

Middleware is specialized software that facilitates communication and data exchange among distributed applications and services \cite{almadani2025publish}. As illustrated in Fig.~\ref{fig:Middleware}, middleware introduces an intermediate software layer that connects applications running on heterogeneous platforms through application programming interfaces (APIs), abstracting underlying operating systems and hardware differences \cite{gambo2025robotics}. This abstraction simplifies development while supporting interoperability and scalability. The Data Distribution Service (DDS) is a middleware standard defined by the Object Management Group (OMG) to support real-time, data-centric communication in distributed systems \cite{gambo2025robotics}. DDS adopts a publish-subscribe model that decouples data producers and consumers and provides a rich set of Quality of Service (QoS) policies to control data delivery behavior. These characteristics have led to its adoption in domains such as robotics, healthcare, and defense, where predictable communication is required.

\IEEEpubidadjcol

A central challenge in remote patient monitoring systems is the integration of heterogeneous devices while ensuring timely and reliable access to patient data. In clinical environments, delays or loss of information can affect decision-making, making communication performance a critical system concern. This work examines a remote patient monitoring architecture built on RTI Connext, which is based on DDS middleware \cite{omgddswikihealthcare2025}, and Internet of Things (IoT) technologies. The objective is to study how middleware-level design and configuration choices influence communication performance under conditions representative of healthcare monitoring scenarios.

Existing monitoring solutions often face difficulties balancing latency, reliability, and scalability in dynamic environments with multiple data sources. Traditional architectures may struggle to sustain high-rate data exchange or to maintain consistent delivery behavior as the number of connected devices increases. These limitations motivate the investigation of middleware-based approaches that provide explicit control over communication properties through configurable QoS policies.

In this work, we design and implement a DDS based real time patient monitoring architecture that enables efficient data exchange among heterogeneous healthcare components. We propose a set of DDS QoS policies and their appropriate configurations for healthcare environments and evaluate their impact on communication efficiency. Furthermore, we introduce a layered data bus architecture that supports modular deployment and provides a scalable foundation for healthcare monitoring systems.

The remainder of the paper is organized as follows. Section~\ref{sec:Literature_Review} reviews related work. Section~\ref{sec:system} describes the proposed system architecture. Section~\ref{sec:experiment} presents the experimental setup. Section~\ref{sec:results} reports and discusses the evaluation results. Section~\ref{sec:conc} concludes the paper and outlines directions for future work.

\section{Literature Review}
\label{sec:Literature_Review}

The rise of chronic diseases, the aging population, and the increasing demand for personalized medical care drive the development of remote healthcare monitoring systems. These systems, powered by IoT technologies, shift healthcare from clinic-centric to patient-centric models, enabling continuous observation of patients outside hospital settings \cite{Peyroteo2021}. Telemedicine platforms and wearable biosensors now support real-time health data acquisition and remote consultation, which is especially important for patients with chronic conditions who require frequent follow-up but face access barriers. Recent reviews of Internet of Medical Things (IoMT) architectures emphasize that non-functional requirements such as latency, reliability, scalability, and security are as critical as sensing accuracy and diagnostic performance \cite{Gallo2025IoMTNonFunctional,Alturki2025IoMT}. These studies underline that communication and middleware choices strongly influence whether telemonitoring systems can safely scale from single-patient prototypes to large multi-patient deployments. Table~\ref{tab:iomt_comparison} summarizes representative recent contributions in IoMT communication, edge computing, and ML-enabled healthcare, grouped by category and focusing on their main contributions and challenges from a communication and QoS perspective.

\subsection{IoT and Remote Healthcare Architectures}

Early IoMT systems focused primarily on integrating wearable sensors, smartphones, and cloud platforms into multi-tier architectures. Kakria et al.\ \cite{kakria2015real} developed a real-time cardiac monitoring system using wearable sensors connected to an Android smartphone, which forwards data and GPS coordinates to a web portal. Their design supports continuous monitoring and alarm generation, but relies on a centralized web server and does not provide fine-grained control over communication QoS. Chaari and Said \cite{ChaariFourati2020} proposed a Remote Healthcare Monitoring System (RHMS) where body sensor nodes communicate with a smartphone coordinator using Bluetooth Low Energy (BLE), and the smartphone relays data to a remote server via Wi-Fi or cellular networks. The system reduces energy consumption at the sensor side but still depends on best-effort IP connectivity without explicit guarantees on delay or loss.

Recent work revisits these architectures with a stronger emphasis on edge processing and real-time constraints. Varma et al.\ \cite{Varma2024IoTEdge} discuss how IoT devices combined with edge computing can support real-time patient monitoring by performing local signal processing close to the data source, thereby lowering end-to-end latency and reducing cloud dependence. Ficili et al.\ \cite{Ficili2025SensorsAI} examine IoT, cloud, and edge computing integration with AI, showing how moving analytics toward the edge improves responsiveness in health-related applications. Uddin and Koo \cite{Uddin2024RemoteMonitoring} review biosensor-based remote monitoring systems built on multi-hop IoT topologies and highlight common bottlenecks related to network congestion and cloud connectivity. These recent studies consistently report that achieving predictable latency and loss characteristics requires explicit QoS management in the communication layer rather than relying solely on best-effort IP or ad hoc gateway implementations.

\subsection{Energy-Efficient and Wearable Sensing Systems}

Energy efficiency is a fundamental challenge in IoT-based healthcare systems because many wearable and implantable devices operate on constrained batteries. Chaari and Said \cite{ChaariFourati2020} adopt BLE to limit power consumption while maintaining acceptable data rates for vital signs. Li et al.\ \cite{Li2023} developed a retractable self-powered respiratory sensor based on a rotating thin-film triboelectric nanogenerator integrated into a wireless monitoring system, showing that self-powered sensing can significantly extend device lifetime. Commercial chest-strap systems such as Resmetrix \cite{resmetrix2025} similarly illustrate how wearable form factors and low-power wireless modules can support long-term respiratory monitoring.

More recent IoMT work combines energy-aware routing, edge computing, and AI-based anomaly detection to manage resource usage at system level. Varma et al.\ \cite{Varma2024IoTEdge} analyze edge-enabled architectures that shift part of the processing workload from cloud servers to intermediate nodes, reducing the volume of raw data transmitted and allowing aggressive duty cycling of sensors. Ibrahim et al.\ \cite{Ibrahim2024SecuritySmartHealthcare} examine security mechanisms for smart healthcare systems and note that cryptographic processing can significantly increase energy consumption on constrained devices, creating a tension between security and battery life. These studies highlight that communication middleware must not only support QoS, but also allow flexible control of message rates and transport policies to balance energy and performance.\cite{gambo2025efficient} demonstrate the effectiveness of DDS security for protecting sensitive information during transmission. 

\subsection{Broker-Based Protocols and Middleware in IoMT}

Broker-based publish-subscribe protocols, particularly Message Queuing Telemetry Transport (MQTT), are widely used to interconnect heterogeneous sensors, gateways, and cloud services in remote health applications. Higher-level frameworks such as MQTT for Sensor Networks (MQTT-SN) and Sparkplug B have been proposed to enhance IoT messaging \cite{kumar2025iot, wang2024cost}. MQTT-SN targets wireless sensor networks with a lightweight, connectionless variant of MQTT that runs over UDP, making it suitable for battery-powered devices. Though MQTT-SN reduces protocol overhead, it still relies on a gateway to an MQTT broker. Sparkplug B is an open specification that layers a uniform data format and state management on top of MQTT for industrial applications. While Sparkplug's schema and birth/death certificates enhance interoperability between devices, it remains dependent on a central MQTT broker. 

Given its application in operational technology (OT) settings, it is worthwhile to compare DDS with other industrial communication standards such as MQTT-Sparkplug and OPC Unified Architecture (OPC UA). Rich information modeling and cross-vendor interoperability are provided by OPC UA, and new additions (such as a pub-sub mode) are intended to satisfy real-time industrial needs. However, in reality, Time-Sensitive Networking or specialized tweaking are frequently required to achieve rigorous low-latency performance with OPC UA \cite{freitas2025opc}. Furthermore, because OPC UA server infrastructure adds layers for data modeling and necessitates security and update management, it may increase maintenance costs in a healthcare setting. 

In contrast, DDS operates brokerlessly, with rich QoS policies that enable peer-to-peer data exchange. As a result, DDS can avoid the single point of failure and latency bottleneck inherent in broker-based approaches. Alshammari et al.\ \cite{11Alshammari2023} employed MQTT with microcontroller-based sensor nodes to relay vital signs to a monitoring server, demonstrating low implementation complexity and adequate latency for single-patient scenarios. Such MQTT-based architectures are attractive because they decouple publishers and subscribers via a broker and reuse well-supported TCP/IP stacks \cite{silvio2019mqtt}. However, they introduce a central point of coordination and potential failure, and their QoS levels are limited to coarse-grained delivery guarantees without built-in notions of bounded latency or jitter. A recent study \cite{kang2020evaluating} comparing DDS vs MQTT in high-frequency IoT workloads shows that DDS often achieves lower latency and more stable throughput.

More recent analyses question the scalability and robustness of purely broker-centric designs. Gallo et al.\ \cite{Gallo2025IoMTNonFunctional} review IoMT systems with a particular focus on non-functional factors and report that many MQTT- or cloud-centric solutions struggle under high device density, especially when encryption and access control are enabled for every message. Pradyumna et al.\ \cite{Pradyumna2024EmpoweringIoMT} survey IoMT architectures with integrated machine learning and identify scalability, interoperability, and security as key challenges, noting that brokered protocols can become bottlenecks as the number of data streams grows. Ibrahim et al.\ \cite{Ibrahim2024SecuritySmartHealthcare} provide a detailed security analysis for smart healthcare systems and emphasize that centralized broker or cloud components widen the attack surface and can hinder resilience in case of failures or attacks. Recent work on machine edge-aware IoT frameworks for health monitoring \cite{Alshuhail2025EdgeAwareIoT} proposes distributed, edge-centric designs that incorporate sensor fusion and AI-driven emergency response, but still relies on application-level mechanisms to manage QoS, leaving transport-level guarantees relatively weak.

Overall, the MQTT/broker-based line of work shows that simple, topic-based communication is sufficient for proof-of-concept systems, but its centralized nature and limited QoS semantics make it difficult to guarantee real-time performance and lossless delivery in dense, multi-patient deployments.

\subsection{DDS-Based Middleware for Real-Time Healthcare Communication}

DDS standard has emerged as a strong candidate middleware for real-time and safety-critical domains because it offers a decentralized, brokerless publish-subscribe model with rich QoS policies \cite{omgddswikihealthcare2025}. Instead of routing messages through a central broker, DDS nodes discover each other and communicate peer-to-peer over a shared “global data space,” which naturally avoids single points of failure and reduces bottlenecks. DDS's applicability for time-sensitive healthcare data transfer has been reinforced by empirical studies, which have demonstrated that it can sustain deterministic, low-latency communication even with OPC UA in Industry 4.0 contexts with non-ideal networks \cite{Ioana2021}. In healthcare settings, DDS can be configured to deliver vital-sign streams with specific reliability, deadline, and history policies, allowing developers to choose between BEST\_EFFORT and RELIABLE modes, define maximum acceptable latency, and control how many samples are stored for late joiners \cite{vectormedicaldds2025,outrightcrmddsiot2025}.

Industrial reports and white papers document DDS deployments in medical environments, including operating rooms, patient monitors, and imaging systems \cite{omgddswikihealthcare2025,vectormedicaldds2025}. These deployments argue that DDS’s fine-grained QoS control and automatic discovery simplify integration of heterogeneous devices while maintaining real-time behavior. However, detailed, open experimental evaluations tailored to healthcare monitoring workloads are still limited compared with the extensive body of work on MQTT- or cloud-based solutions. Existing DDS evaluations often target robotics or generic cyber-physical systems rather than patient monitoring workloads, and they rarely compare DDS against lighter protocols under varying message sizes, numbers of subscribers, and QoS configurations.

This gap motivates experimental studies that characterize DDS performance for healthcare monitoring traffic. In particular, there is a need to quantify how DDS latency, throughput, and packet loss behave when switching between RELIABLE and BEST\_EFFORT QoS profiles and when scaling the number of publishers and subscribers in realistic hospital network scenarios.

\subsection{Machine Learning and AI-Enabled IoMT Architectures}

Parallel to advances in communication middleware, there has been rapid growth in Machine Learning (ML) and AI techniques for analyzing physiological signals, detecting anomalies, and predicting adverse events in IoMT systems \cite{11157822, 10578130}. Surveys such as Gallo et al.\ \cite{Gallo2025IoMTNonFunctional} and Pradyumna et al.\ \cite{Pradyumna2024EmpoweringIoMT} emphasize that recent IoMT solutions increasingly embed ML models either at the edge or in the cloud to enable early detection of deterioration, risk stratification, and personalized feedback. Jafari and Adibnia \cite{Jafari2025FLBigchainDB} propose an architecture that combines federated learning, BigchainDB, and blockchain technology to secure health data while collaboratively training ML models, reducing the need to transmit raw data and improving robustness against attacks.

Federated learning and split learning have been explored as privacy-preserving training paradigms for wearable and IoMT devices. Ni et al.\ \cite{Ni2024FedSL} introduce FedSL, a federated split learning framework that enables multiple resource-constrained IoMT devices to jointly train deep models for healthcare analytics without sharing raw data, while offloading part of the model to more capable nodes. Albogamy et al.\ \cite{Albogamy2025FedHAR} develop a federated-learning-based human activity recognition system using a hybrid LSTM-GRU architecture deployed across IoMT devices to classify activities from wearable sensors. These works highlight that distributed ML workloads are highly sensitive to communication latency, jitter, and loss because model updates must be exchanged frequently between many devices and aggregators.

Recent surveys also emphasize the tight coupling between IoMT communication, ML integration, and security. Pradyumna et al.\ \cite{Pradyumna2024EmpoweringIoMT} discuss how ML enhances anomaly detection and decision support, but also stress that unreliable communication or packet loss can degrade model performance and delay responses. Alturki et al.\ \cite{Alturki2025IoMT} map open challenges around interoperability, security, and latency in IoMT, pointing out that many ML-enhanced architectures still rely on generic IP or broker-based messaging without explicit real-time guarantees.

Although the present work does not implement any ML functionality, the proposed DDS-based communication model is directly compatible with such AI-enabled frameworks. By providing configurable QoS policies for reliability, latency, and resource usage in a decentralized fashion, DDS can supply high-quality, lossless data streams to downstream ML modules or federated learning agents. Conversely, ML-based controllers could exploit DDS monitoring statistics to adapt QoS parameters dynamically, prioritize critical topics, or schedule bandwidth among competing data flows. In this sense, a carefully characterized DDS communication layer, as investigated in this paper, can serve as a robust substrate for future integration of AI-based analytics and optimization in IoMT healthcare systems.

\begin{table*}[!ht]
\centering
\caption{Representative IoMT Communication and Analytics Approaches}
\label{tab:iomt_comparison}
\renewcommand{\arraystretch}{1.2}
\begin{tabular}{p{2.0cm} p{2.0cm} p{2.3cm} p{4.5cm} p{4.5cm}}
\toprule
Ref / Year & Category & Technology / Middleware & Main Contribution & Challenges \\
\midrule
\cite{Alturki2025IoMT} / 2025 &
Survey / IoMT &
IoMT architectures and enabling technologies &
Maps the IoMT landscape with emphasis on current challenges and future research trends in connectivity, security, and interoperability for medical devices. \newline &
Limited emphasis on quantitative QoS evaluation and detailed comparison of specific middleware stacks under realistic monitoring workloads. \\

\cite{Gallo2025IoMTNonFunctional} / 2025 &
Survey / Systems &
IoMT platforms and services &
Systematic review of IoMT systems focusing on non-functional requirements such as scalability, latency, security, and reliability. &
Most studies reviewed rely on cloud or broker-based communication; detailed assessment of decentralized RT middleware like DDS is largely absent. \newline \\

\cite{Ficili2025SensorsAI} / 2025 &
Edge / AI integration &
IoT-cloud-edge with AI &
Discusses architectures that move AI-driven analytics from cloud to edge for time-sensitive applications, including health-related monitoring. &
Does not provide low-level measurements of latency and packet loss for specific middleware; assumes generic IP and message brokers. \newline \\

\cite{Alshuhail2025EdgeAwareIoT} / 2025 &
Edge framework &
Edge-aware IoT for health monitoring &
Proposes a machine edge-aware IoT framework with sensor fusion and AI-driven emergency response in decentralized networks. &
Relies on application-level orchestration; communication layer remains best-effort and may struggle to provide strict latency bounds in dense deployments. \newline \\

\cite{Albogamy2025FedHAR} / 2025 &
ML / HAR &
Federated learning with hybrid LSTM-GRU on IoMT devices &
Introduces a federated learning architecture for human activity recognition that preserves privacy while exploiting distributed wearable data. &
Learning performance and convergence depend on timely, reliable exchange of model updates; communication substrate is not explicitly optimized or characterized. \newline \\

\cite{Pradyumna2024EmpoweringIoMT} / 2024 &
Survey / ML-enabled IoMT &
IoMT with ML integration &
Reviews IoMT systems with integrated ML, highlighting evolution, ML integration patterns, and security and interoperability challenges. &
Identifies communication and QoS issues but leaves open how to realize fine-grained QoS control and real-time guarantees in practice. \newline \\

\cite{Varma2024IoTEdge} / 2024 &
Edge-enabled monitoring &
IoT and edge computing for real-time patient monitoring &
Analyzes how edge computing and IoT can support real-time medical monitoring and predictive analytics while reducing cloud dependence. &
Focuses on architectural patterns; empirical analysis of transport-level QoS and comparative middleware performance is limited. \newline \\

\cite{Ibrahim2024SecuritySmartHealthcare} / 2024 &
Security / systems &
Smart healthcare systems &
Provides a security analysis of smart healthcare systems and discusses threats and countermeasures for connected medical devices. &
Security mechanisms introduce additional overhead; the trade-off between strong security and real-time QoS is not systematically quantified. \newline \\

\cite{Ni2024FedSL} / 2024 &
Federated / split learning &
FedSL on wearable IoMT devices &
Proposes a federated split learning framework that enables collaborative healthcare analytics on resource-constrained wearable IoMT devices. &
Assumes reliable connectivity for exchanging intermediate activations and model updates; impact of network latency and loss on training dynamics is not experimentally explored. \newline \\

\cite{Jafari2025FLBigchainDB} / 2025 &
Security + FL &
Federated learning with BigchainDB and blockchain &
Designs a secure IoMT architecture that combines federated learning, BigchainDB, and blockchain to protect health data and improve ML robustness. &
Communication patterns become more complex and latency-sensitive due to blockchain and FL coordination; middleware is not specialized for real-time QoS. \newline \\

\textbf{This work} / 2025 &
DDS-based healthcare communication &
DDS middleware and TCP sockets &
Experimental comparison of DDS QoS profiles against socket-based communication for healthcare monitoring traffic, including latency, throughput, and packet loss under varying payloads and network setups. &
Focuses on communication-layer metrics; future work can integrate ML modules on top of the characterized DDS data streams and study closed-loop AI-enabled control. \\
\bottomrule
\end{tabular}
\end{table*}

\section{Proposed System}
\label{sec:system}

\subsection{DDS Middleware Architecture}

The real-time patient monitoring system uses DDS middleware to enable communication among distributed medical devices, clinical applications, and database systems. DDS supports a data-centric publish-subscribe model in which communication occurs through named {Topics} over a shared {Databus}. Participants in this model are represented as {DomainParticipants}, which define the scope of interaction within a specific domain. Each {DomainParticipant} may include one or more {Publishers} and {Subscribers}. Data transmission is performed by {DataWriters} and {DataReaders} associated with specific {Topics}.

In the current system design, all components, including patient-side sensors, nurse and doctor interfaces, and the database application, are implemented as DDS DomainParticipants. These participants are grouped across two domains: Domain 10, which represents the patient room, and Domain 20, which includes ward-level functions such as data archival, clinical monitoring, and administrative control. A Routing Service connects these domains while preserving DDS semantics. The structure of the system is shown in Fig~\ref{fig:system-architecture}, where all key components and communication paths are labeled according to DDS terminology.

\subsection{Quality of Service (QoS) Management}

The DDS middleware allows developers to assign QoS policies to all communication paths. These policies define the required characteristics of data exchange, including reliability, latency constraints, resource limits, and data persistence. For this system, the Reliable QoS policy is used in the primary implementation. This policy ensures that every data sample transmitted by a DataWriter is delivered to all matched DataReaders, with retransmissions occurring when delivery is not acknowledged.

In parallel to this configuration, the system’s performance was also evaluated using the Best Effort QoS policy. Under Best Effort, data samples are transmitted without requiring acknowledgment or retransmission, which reduces delivery latency but allows for possible data loss. Both configurations, Reliable and Best Effort, were applied to representative communication flows in the system. Their impact on system behavior was measured in terms of latency, throughput, and data completeness.

The comparison revealed that the Reliable QoS configuration provides full delivery assurance, which is necessary for topics involving patient vitals or medical decision support, such as SensorInfo and DBStore. However, it introduces additional latency due to acknowledgment and retransmission mechanisms. In contrast, the Best Effort configuration achieves lower latency and higher throughput in scenarios with non-critical data, such as frequent environmental readings or monitoring logs, though it does not guarantee delivery.

This comparative use of both QoS policies demonstrates the trade-offs involved in designing healthcare monitoring systems. The application of each QoS mode is determined based on the criticality of the data and the operational demands of the system. Topics such as SensorInfo, DBStore, ManagerInfo, and DBQuery are configured with QoS policies according to their role in the system, as indicated in Table~\ref {tab:qos_application_roles}.

\subsection{Patient Room Domain}

Domain 10, shown in the right side of Fig~\ref{fig:system-architecture}, encompasses the patient room environment. This domain includes multiple Vital Sensors that measure physiological signals such as heart rate, blood pressure, and body temperature. Sensor readings are first processed by a dedicated Sensor Data Fusion module, which performs data aggregation and formatting.

Processed sensor data is published to the local databus by a Publisher within the SensorApp DomainParticipant. The publishing is performed by a DataWriter that transmits messages on the SensorInfo topic. Communication occurs over the Patient Room Databus, which supports intra-room messaging among local devices.

The SensorInfo topic is configured with Reliable QoS in the system’s primary setup to ensure that all transmitted data reaches subscribing applications in Domain 20. For comparative evaluation, the same topic was tested under Best Effort QoS. In this configuration, the system achieved faster transmission times but occasionally missed updates during network contention or congestion. These results confirm the sensitivity of medical data flows to reliability policies and support the selection of Reliable QoS for mission-critical data streams.

\subsection{Ward-Level Domain}

\begin{figure*}[!t]
  \centering
  \includegraphics[scale=0.5]{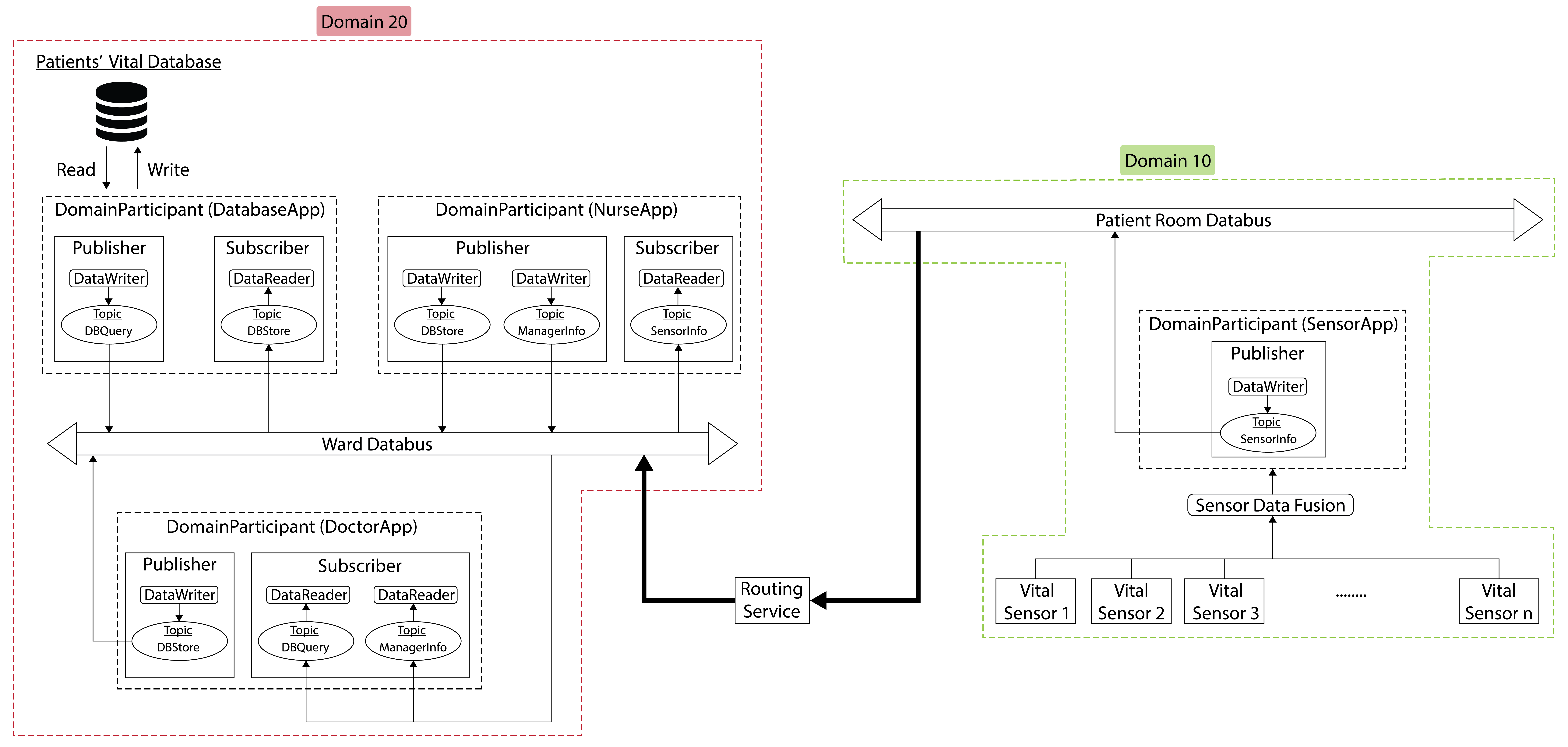}
  \caption{System architecture showing DDS Domains 10 and 20, domain participants, databuses, and topic-level communication paths.}
  \label{fig:system-architecture}
\end{figure*}

Domain 20, as shown in the left side of Fig~\ref{fig:system-architecture}, includes clinical and administrative applications. Three major DomainParticipants operate in this domain: DatabaseApp, NurseApp, and DoctorApp.

The DatabaseApp handles the archival and retrieval of patient data. Its Publisher writes to the DBStore topic, and its Subscriber queries stored records via the DBQuery topic. These topics use Reliable QoS to ensure reliable retrieval and storage of patient data.

The NurseApp includes DataWriters for the DBStore and ManagerInfo topics and a DataReader for the SensorInfo topic. This configuration allows nurses to receive real-time sensor updates and issue data or administrative commands. The reliability of SensorInfo is necessary to ensure timely clinical responses.

The DoctorApp includes similar functionality, subscribing to ManagerInfo and DBQuery, and publishing updates to DBStore. This allows physicians to access and update clinical records based on real-time data and stored patient histories.

All Domain 20 communication is performed over the Ward Databus. Data from Domain 10 is propagated to Domain 20 through the Routing Service, which manages topic mapping and QoS translation between the two domains. 

\begin{table*}[!t]
\centering
\caption{Configured QoS Policies}
\label{tab:qos_application_roles}
\renewcommand{\arraystretch}{1.2}
\begin{tabular}{l l l l l l p{5.5cm}}
\toprule
\textbf{Application} &
\textbf{Reliability} &
\textbf{History} &
\textbf{Durability} &
\textbf{Deadline} &
\textbf{Liveliness} &
\textbf{Resource Limits} \\
\midrule
Sensor &
RELIABLE &
KEEP\_ALL &
VOLATILE &
10 s &
Automatic &
max\_instances = 1024, \newline max\_samples\_per\_instance = 100, \newline max\_samples = 10000 \newline \\

Nurse Station &
RELIABLE &
KEEP\_ALL &
VOLATILE &
11 s &
Automatic &
max\_instances = 1024, \newline max\_samples\_per\_instance = 100, \newline  max\_samples = 10000 \newline \\

Doctor &
RELIABLE &
KEEP\_ALL &
VOLATILE &
Infinite &
Automatic &
max\_instances = 1024, \newline max\_samples\_per\_instance = 100, \newline  max\_samples = 10000 \newline \\

Database &
RELIABLE &
KEEP\_ALL &
VOLATILE &
Infinite &
Automatic &
max\_instances = 1024, \newline max\_samples\_per\_instance = 100, \newline  max\_samples = 10000 \\
\bottomrule
\end{tabular}
\end{table*}

\section{Experimentation}
\label{sec:experiment}

To evaluate the proposed system, we implemented a functional prototype that simulates a real-time patient monitoring scenario in a healthcare environment. All experiments and prototype implementation were conducted using RTI Connext DDS version 7.5.0. The experimental setup consists of three main components: the SensorApplication, the NurseApplication, and the DoctorApplication. The SensorApplication emulates the behavior of a medical sensor installed in a hospital ward. It continuously monitors patient vitals and publishes data to the Patient Room Data Bus. The data includes patient identification, patient name, room number, sensor identification, blood pressure, heart rate, and oxygen saturation. This simulates a realistic environment in which physiological data is continuously gathered from patients.

\begin{figure}[htbp]
    \centering
       \includegraphics[width=0.9\linewidth]{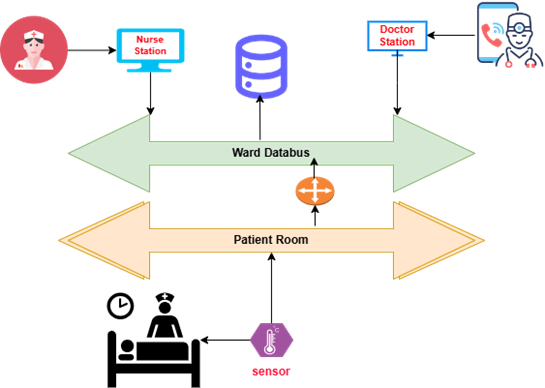}
    \caption{Experimental architecture}
    \label{fig:experimental-architecture}
\end{figure}

The NurseApplication subscribes to this data and performs several functions. It stores the received data in a My Structured Query Language (MySQL) database, evaluates the measurements to identify abnormal readings, and generates alerts if critical thresholds are exceeded. Additionally, the NurseApplication responds to requests from the DoctorApplication by providing relevant patient data. The DoctorApplication displays alerts received from the NurseApplication, enabling timely clinical review. It also allows medical personnel to send commands and retrieve information about active sensors and patient records stored in the system. All communication between components is facilitated through a Data Distribution Service, which ensures reliable and scalable data exchange. The overall implementation architecture used in the experiment is illustrated in Fig~\ref{fig:experimental-architecture}.

In our experimental setup, we applied a set of DDS QoS policies to ensure reliable communication between distributed heterogeneous components of the system. The {RELIABILITY} QoS policy was configured as Reliable, which guarantees that published data is delivered to all matched subscribers. This setting is particularly important in our healthcare context, where it is critical to ensure that vital patient readings, such as heart rate and oxygen saturation, reach their intended consumers, such as the nurse or doctor applications, without loss.

To manage memory for reliability, we also configured the {RESOURCE\_LIMITS} and {HISTORY} QoS policies. In addition, the {LIVELINESS} policy was configured to detect the presence of active participants in the system, as shown in Table~\ref {tab:qos_application_roles}.

Before conducting the distributed  experiments, we first evaluated the system performance on a single host using the shared-memory (SHMEM) transport. The hardware platform consisted of an Intel® Pentium® CPU 2020M @ 2.40 GHz (dual-core), 12 GB RAM, and a Realtek PCIe FE network interface (100 Mbps Fast Ethernet). In this configuration, the NIC was not part of the measurement path; all DDS communication was carried over shared memory. Table \ref{tab:system_configuration} shows the items used in the distributed experiments. The experiments were conducted using two Dell OptiPlex machines running Windows 10 Pro (64-bit), each equipped with an Intel Core i5 processor (6 cores, 3.10 GHz) and 8 GB of DDR4 RAM. The nodes were connected via a 1 Gigabit Ethernet network through a Linksys ST2800 Gigabit switch. These settings and configurations allowed us to evaluate the proposed system under realistic conditions, reflecting practical requirements for real-time, publish-subscribe communication in healthcare monitoring environments.

\begin{table}[!ht]
\centering
\caption{System Configuration}
\label{tab:system_configuration}
\renewcommand{\arraystretch}{1.2}
\begin{tabular}{l l}
\toprule
\textbf{Item} & \textbf{Description} \\
\midrule
Machines & Dell OptiPlex 5080 \\
Os & Windows 10 Pro (64-bit) \\
CPU & Intel Core i5 (6 cores, 3.10 GHz) \\
RAM & 8 GB DDR4 on each host \\
NIC & 1 Gbps Ethernet \\
Switch & Linksys ST2800 Gigabit switch \\
\bottomrule
\end{tabular}
\end{table}

\section{Result and Discussion}
\label{sec:results}

In this section, we present and analyze the performance results of the proposed system. The primary objective of our evaluation is to measure communication efficiency and reliability under different middleware configurations, comparing DDS-based messaging with conventional socket-based communication. Experiments were conducted using both Reliable and Best Effort QoS modes in DDS to highlight the trade-offs between delivery assurance and transmission efficiency. The section also compares the performance of the DDS modes (i.e., both Reliable and Best Effort QoS modes) with socket communication. Performance metrics considered include latency, throughput, and packet loss, which collectively capture the responsiveness and robustness of the system. The metrics were obtained using  Perftest \cite{bode2023systematic}. This section aims to demonstrate the advantages of a decentralized publish-subscribe approach over traditional point-to-point communication in real-time healthcare monitoring scenarios.

\subsection{Single Machine}
\label{subsec:single_machine_result}
To evaluate the minimum latency and maximum achievable throughput, we first conducted experiments on a single machine. This local evaluation isolates the core messaging behavior of the system without the influence of network delays or distributed overhead. The section presents the performance of this experiment, presenting results for latency, throughput, and packet loss under both Reliable and Best Effort QoS configurations. The results provide a baseline characterization of DDS messaging efficiency, showing how the additional reliability mechanisms in the Reliable QoS setting influence performance as the message size increases. These single-machine measurements are essential for distinguishing protocol-level behavior from network-induced effects, and they serve as a reference point for later distributed evaluations.



\begin{figure}[!ht]
    \centering
    \subfloat[Latency comparison]{
        \includegraphics[width=0.23\textwidth]{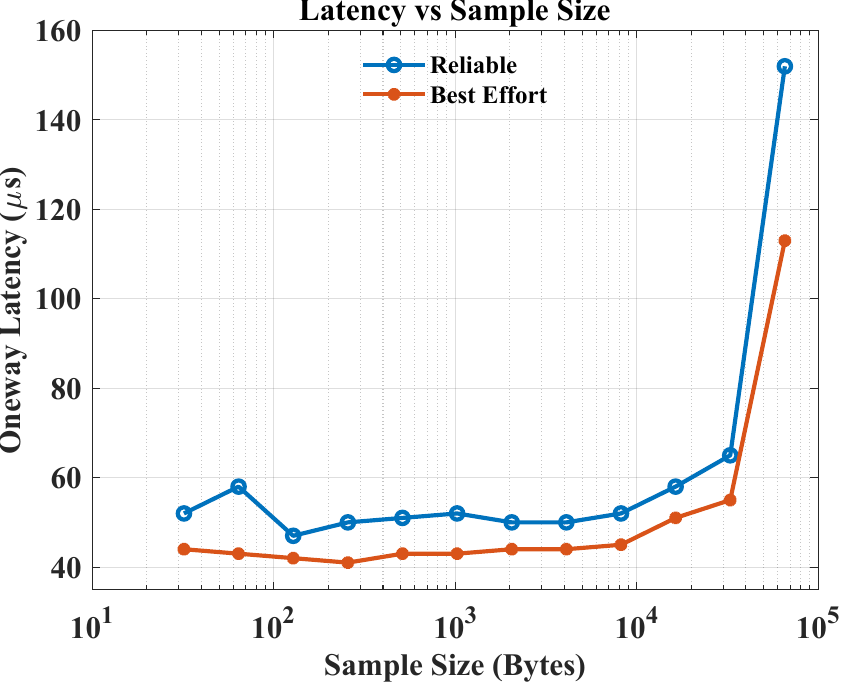}
        \label{fig:latency-comparison_single_machine}
    }\hfill
    \subfloat[Throughput comparison]{
        \includegraphics[width=0.23\textwidth]{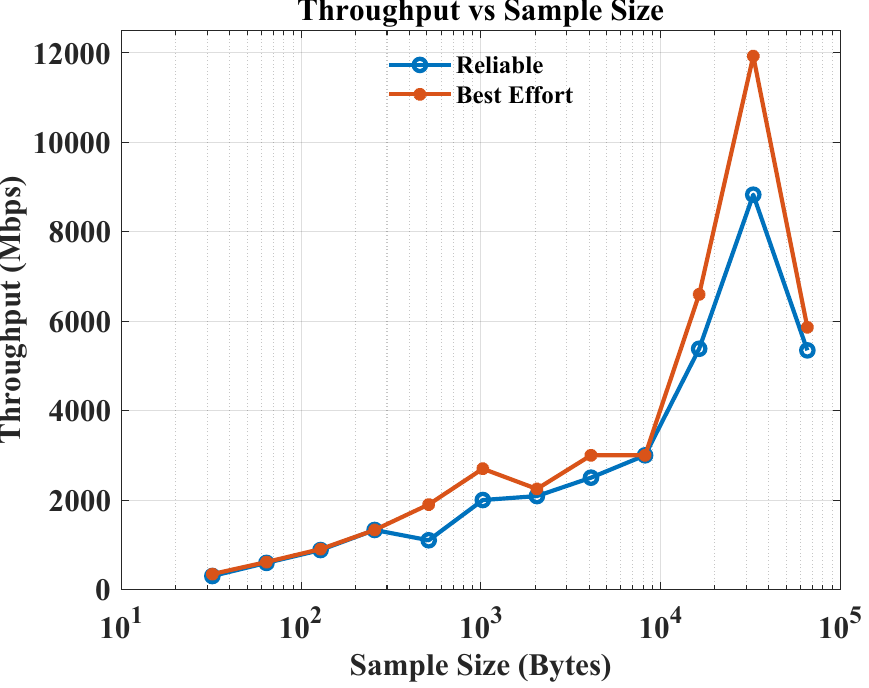}
        \label{fig:Througput-comparison_single_machine}
    }
    \caption{Latency and throughput comparison of DDS on a single machine}
    \label{fig:latency-throughput_single_machine}
\end{figure}

Fig~\ref{fig:latency-comparison_single_machine} shows the latency results for Reliability and Best Effort QoS policies with different sample sizes. For smaller sample sizes (32 to 4096 bytes), both QoS policies exhibit similar latency, though Reliability generally has slightly higher latency due to its acknowledgment mechanism. This latency remains consistent within the range of 40-60~\textmu s for small messages. However, as the sample size increases (8192 bytes and above), latency starts to increase more sharply, especially for the Reliability QoS. For the largest sample size (65536 bytes), the latency for Reliability rises significantly, while Best Effort latency also increases but remains lower throughout. This suggests that the retransmission and delivery guarantees in Reliability become more noticeable as the payload size increases.



Fig.~\ref{fig:Througput-comparison_single_machine} shows the throughput across different payload sizes for both Reliable and Best Effort QoS modes. As the sample size increases, throughput rises for both policies. For smaller payloads (up to 4096 bytes), the throughput is almost identical, but for larger payloads (16,384 bytes and above), Best Effort achieves higher throughput. This advantage comes from the absence of additional reliability mechanisms, which in the Reliable mode add processing overhead to guarantee data delivery. Best Effort reaches a peak throughput exceeding 10,000 Mbps, while Reliable remains slightly lower. At very large sample sizes, a gradual decline in throughput is observed due to RTI transport configuration limits, particularly the Shared Memory (SHMEM) transport's {message\_size\_max} of 65,336 bytes and the handling of unbounded sequence types, which trigger asynchronous publishing and increase processing overhead.


It was observed that there was no packet loss under either Reliable or Best Effort QoS configurations across all tested payload sizes. This demonstrates that DDS can provide dependable data delivery even under high message loads, making it suitable for healthcare monitoring applications where lossless transmission of patient data is essential. The absence of packet loss is expected because the experiment was conducted on a single machine, with communication managed through shared memory since all components were colocated. To further evaluate performance under realistic network conditions, a second experiment was carried out using two machines, one acting as the publisher and the other as the subscriber, as described in Section~\ref{subsec:dds_comm}.



\subsection{DDS vs Socket Communication}
\label{subsec:dds_comm}
To further evaluate the efficiency of the proposed DDS-based architecture, we compared its performance against conventional socket-based communication. While plain sockets offer minimal transmission overhead, they lack essential features such as discovery, reliability control, and configurable QoS policies. However, DDS provides a decentralized publish-subscribe model with tunable QoS that can better support heterogeneous and real-time healthcare applications. In this subsection, we analyze latency, throughput, and packet loss across both approaches to highlight the trade-offs between raw transmission speed and dependable, scalable communication.

\begin{figure}[!ht]
    \subfloat[Average]{
    \includegraphics[width=0.23\textwidth]{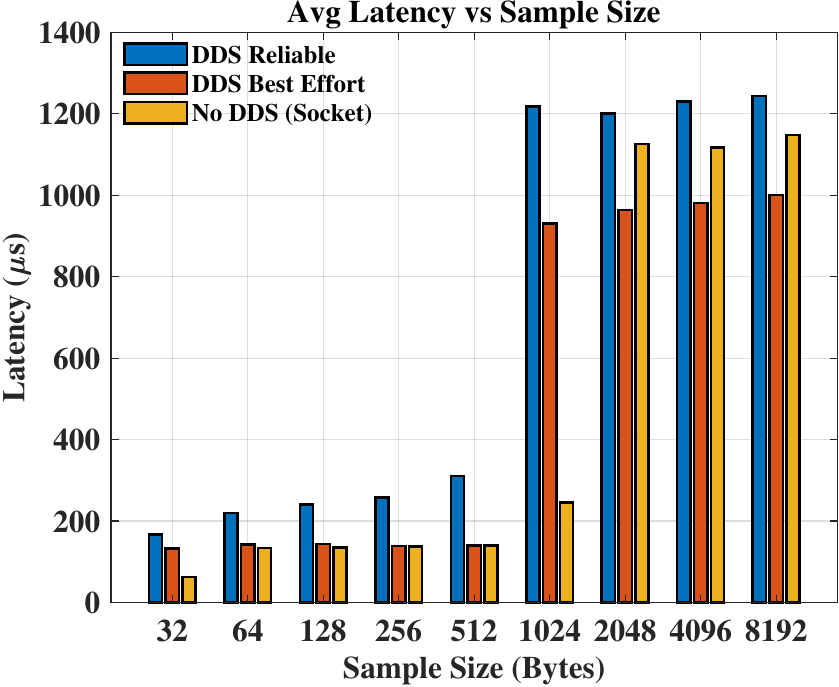}
    \label{fig:avg_latency}
    }\hfill
    \subfloat[Minimum]{
    \includegraphics[width=0.23\textwidth]{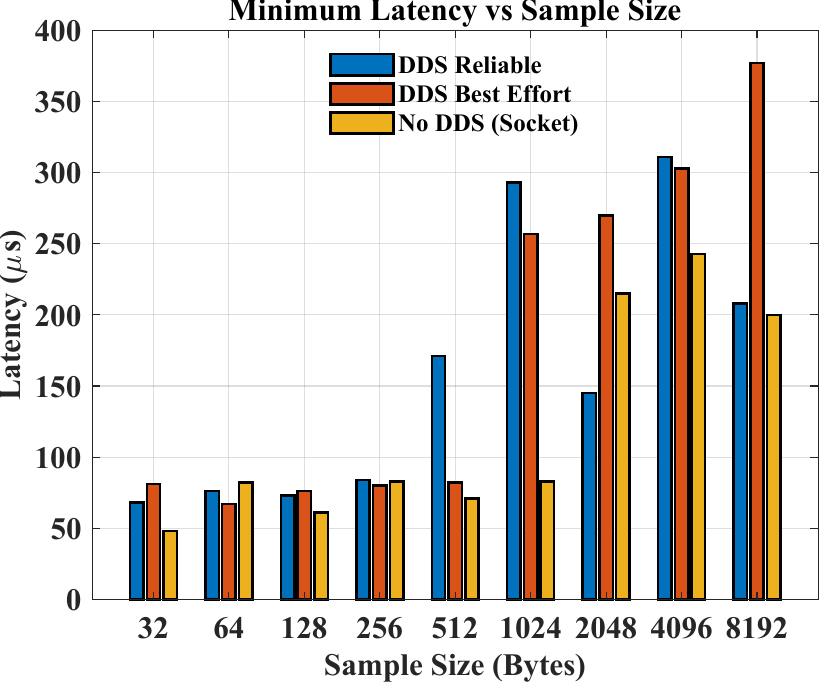}
    \label{fig:min_latency}
    }\\
    \subfloat[Maximum]{
    \includegraphics[width=0.23\textwidth]{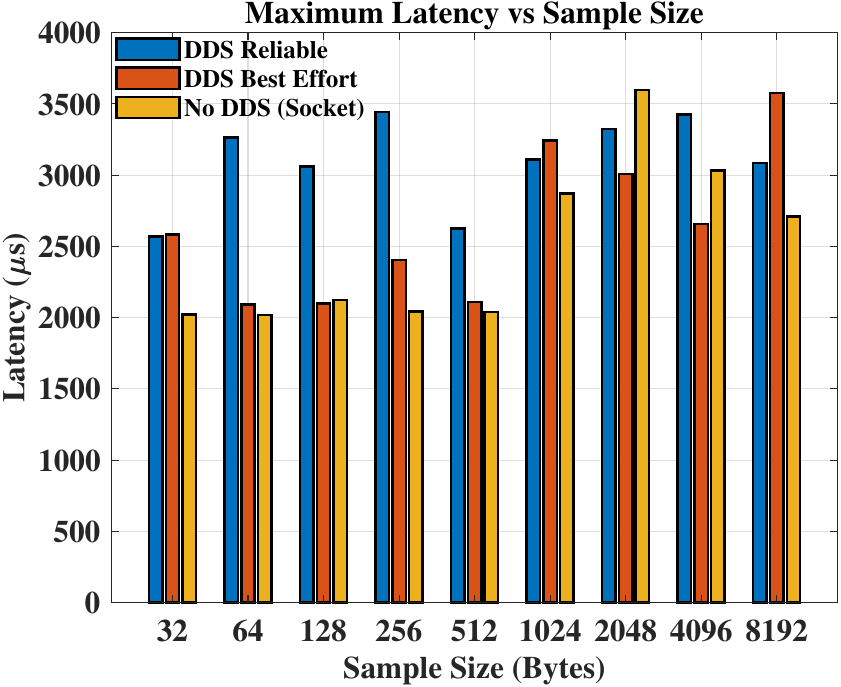}
    \label{fig:max_latency}
    }\hfill
    \subfloat[Standard Deviation]{
    \includegraphics[width=0.23\textwidth]{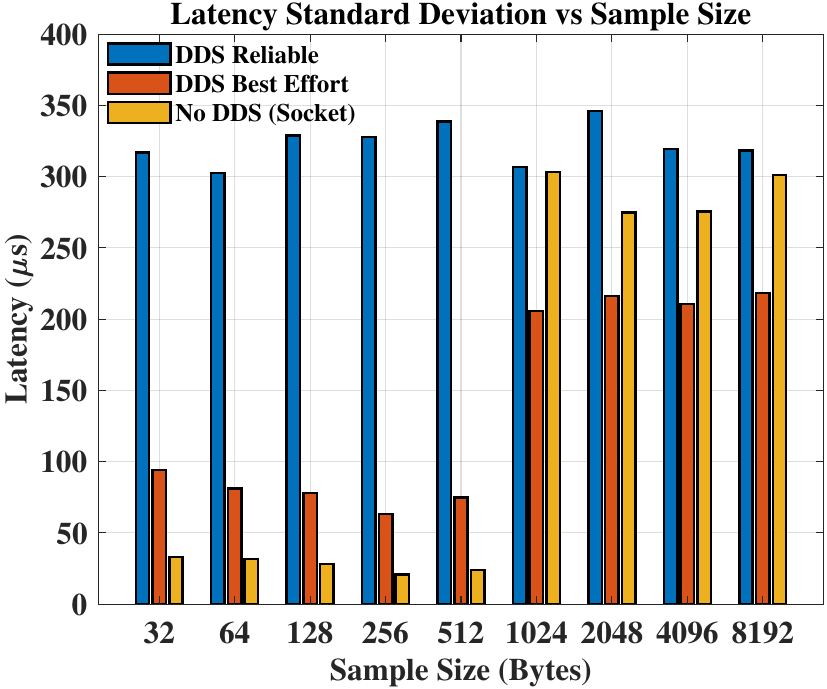}
    \label{fig:std_latency}
    }
    \caption{Latency of Proposed System using DDS and Socket}
    \label{fig:latency-comparison}
\end{figure}

The latency results in Fig.~\ref{fig:latency-comparison} highlight the differences between DDS-based communication and conventional sockets across multiple dimensions. Fig.~\ref{fig:avg_latency} shows that average latency increases gradually with payload size for both DDS and socket communication, with sockets remaining lower overall for small messages (32 to 1024~B) due to DDS middleware overhead. However, for larger packets ($\geq$ 2048~B), Best Effort QoS achieves lower latency than sockets, while Reliable QoS exhibits higher latency. This behavior reflects the trade-off between the two DDS modes: Reliable QoS ensures guaranteed delivery at the cost of increased latency, whereas Best Effort QoS provides lower latency and higher throughput but without delivery guarantees.  

Fig.~\ref{fig:min_latency} indicates that sockets achieve the lowest minimum latency, particularly at smaller payloads. The figure also shows that Reliable QoS occasionally provides lower latency for small packets, while Best Effort QoS sometimes incurs higher latency when transmitting larger packets. Similarly, Fig.~\ref{fig:max_latency} demonstrates that maximum latency is generally higher for DDS under the Reliable QoS mode, but remains below 3,500~\textmu s across all message sizes. In contrast, sockets and Best Effort QoS occasionally exceed 3{,}500~\textmu s for larger packets. 

The standard deviation (SD) of latency in Fig.~\ref{fig:std_latency} can be interpreted as a measure of jitter, since it reflects the variability in packet delay across multiple transmissions~\cite{rao2009every}. Jitter refers to inconsistency in packet latency, which can cause issues such as choppy audio/video, dropped calls, interrupted online gaming, or delayed responses in surgical tele-robotics. The figure shows that the SD of Reliable QoS remains steady across all packet sizes, albeit higher than the others. The benefit of this behavior is predictability: consistent jitter ensures latency remains bounded and easier to compensate for in real-time healthcare applications, such as ICU ECG monitoring, ventilator supervision, or wearable glucose tracking, where dependable performance is more critical than occasional low-latency outliers that could still allow harmful delays. In addition, Best Effort QoS outperforms socket communication at larger packet sizes, whereas sockets are faster for smaller packets; this crossover occurs because the fixed middleware overhead of DDS has less relative impact as payload size increases.

Therefore Fig.~\ref{fig:latency-comparison} shows that the advantage of plain sockets is only noticeable when the data length is less than 1024 bytes. Unlike DDS, plain socket-based communication provides only basic data transfer. There is no automatic peer discovery, connections must be created and maintained manually, and there is no built-in mechanism to guarantee delivery or recover lost messages. Furthermore, socket communication does not have configurable quality-of-service policies for latency, reliability, or prioritization. DDS, on the other hand, provides discovery mechanism, and fine-grained QoS management, which are particularly valuable in continuous patient-monitoring scenarios where timely and dependable data delivery is critical.

\begin{figure}[!ht]
    \subfloat[Total Samples]{
    \includegraphics[width=0.23\textwidth]{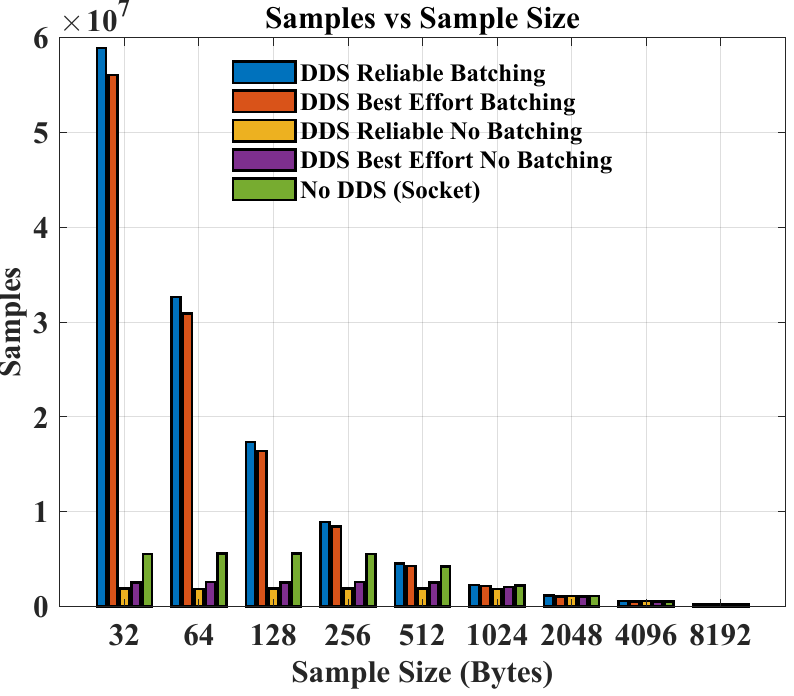}
    \label{fig:total_samples}
    }\hfill
    \subfloat[Average Samples]{
    \includegraphics[width=0.23\textwidth]{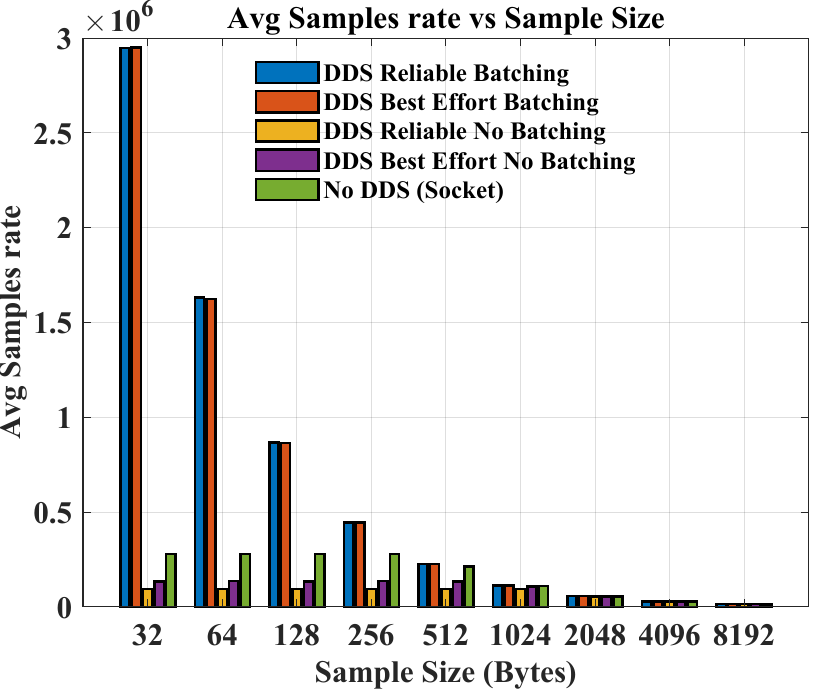}
    \label{fig:ave_sample}
    }\\
    \subfloat[Lost Samples]{
    \includegraphics[width=0.23\textwidth]{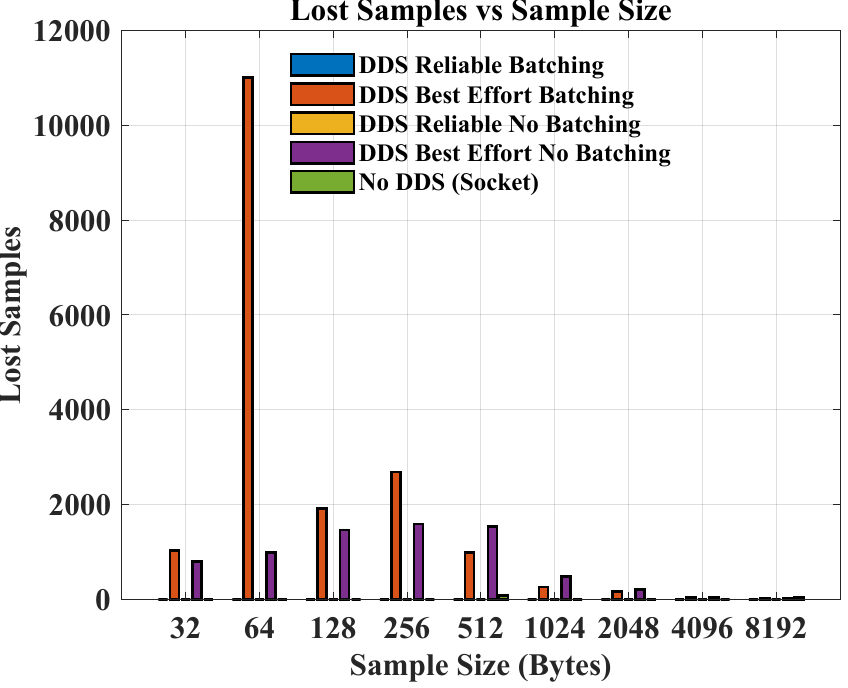}
    \label{fig:lost_samples}
    }\hfill
    \subfloat[Average Throughput]{
    \includegraphics[width=0.23\textwidth]{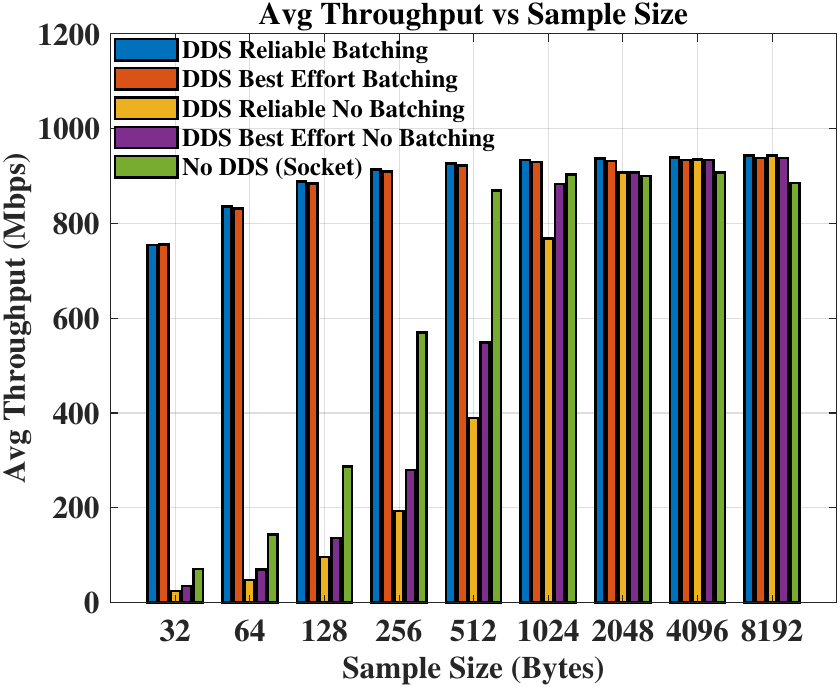}
    \label{fig:avg_throughput}
    }
    \caption{Other performance metrics of the proposed system using DDS and Socket communication}
    \label{fig:throughput-comparison}
\end{figure}

The additional performance metrics in Fig.~\ref{fig:throughput-comparison} provide further insights into throughput, packet loss, and the impact of batching (using the Batch QoS policy) on DDS communication compared to sockets. In RTI Connext DDS, the Batch QoS (BATCH) policy allows multiple data samples to be aggregated into a single network packet, thereby increasing effective throughput by reducing the per-packet overhead associated with transmitting many small messages~\cite{RTI2021batching}. This mechanism improves network utilization, particularly for small data payloads, by sending larger, batched packets, which can also reduce Real-Time Publish-Subscribe (RTPS) traffic for Reliable topics. The presented results were obtained using Perftest, executed for 20~s with one machine acting as the publisher and another as the subscriber.

Fig.~\ref{fig:total_samples} shows the total number of transmitted samples across different payload sizes. DDS with batching, under both Reliable and {Best Effort} modes, consistently achieves the highest sample counts, significantly outperforming both sockets and DDS without batching. This demonstrates the efficiency of batching in aggregating small messages into larger ones, thereby reducing transmission overhead. The result agrees with the RTI Connext documentation, which notes that batching is most effective for packets smaller than 2,048~B~\cite{RTI2022batch}. The figure also shows that socket communication transmits more packets than Reliable and {Best Effort} without batching at small payloads, since middleware overhead increases transmission latency. However, for larger packets ($\geq$ 2048~B), socket performance degrades, allowing the non-batched DDS configurations to catch up. Another noteworthy observation for small packets is that Reliable with batching outperforms Best Effort with batching, while the opposite trend is observed without batching. This is intuitive because losing a batched transmission has a higher cost, making it advantageous to ensure reliability when batching is enabled.

Similarly, Fig.~\ref{fig:ave_sample} illustrates the average number of samples transmitted per second, where DDS with batching sustains higher throughput across all payload sizes. The results also show that Reliable with batching and Best Effort with batching achieve nearly identical average sample rates. This supports the earlier observation that Reliable with batching attains a higher total number of samples (Fig.~\ref{fig:total_samples}) because it experiences fewer losses. Avoiding the overhead associated with retransmitting large batched packets and additional packing time allows the Reliable mode to maintain efficiency, whereas packet loss under Best Effort results in a larger performance penalty due to the loss of entire batches.

To further investigate the poor performance of Best Effort with and without batching, we examined the number of lost samples. Fig.~\ref{fig:lost_samples} presents the loss distribution across payload sizes. The results show no packet loss for Reliable communication, with or without batching, since DDS guarantees reliable delivery. Socket communication also shows very low packet loss, with only 83 and 49 samples lost at 512~B and 8192~B, respectively. This low loss rate is expected, as sockets avoid the middleware overhead of DDS. In contrast, and in line with our earlier observations, the Best Effort modes exhibit significant packet loss in both batching and non-batching configurations. At smaller payload sizes ($\leq$ 256~B), batching suffers higher losses because each batch contains a large number of aggregated samples, amplifying the impact of a single dropped packet. Conversely, at larger payloads, the non-batching mode loses more samples, since fewer packets are combined per transmission, reducing batching efficiency. These results confirm that the lack of delivery guarantees, coupled with middleware overhead, causes Best Effort QoS to perform the poorest in terms of lost samples.

Fig.~\ref{fig:avg_throughput} compares the average throughput. The results show that batching (both Reliable and Best Effort) achieves the highest throughput across all payload sizes, with performance improving as sample size increases until stabilizing around 512~B. Reliable with batching slightly outperforms Best Effort with batching, since no packets are lost in the Reliable mode. The figure also shows that Reliable without batching underperforms compared to Best Effort without batching, particularly at smaller payload sizes where batching has the greatest impact. This confirms that batching compensates for the overhead introduced by the Reliable QoS. Socket communication demonstrates reasonable performance for small payloads, outperforming DDS without batching, but it lags behind DDS with batching. At larger payload sizes ($\geq$ 2048~B), sockets underperform all DDS configurations, as DDS manages large packet transmissions more efficiently. These results highlight the critical role of batching in improving network performance and ensuring scalability.



Therefore, the BATCH QoS policy enables DDS to significantly improve throughput by aggregating smaller messages into larger payloads, thereby reducing transmission overhead. These results confirm that while sockets provide low-overhead communication for small messages, DDS with batching offers superior throughput scalability, bandwidth efficiency, and reliability across a wide range of payload sizes. This makes batching particularly effective for high-frequency, small-payload communication in real-time systems such as healthcare monitoring, where many embedded devices and sensors publish frequent small packets of patient data.



\subsection{Performance over Wireless Communication}
\label{subsec:wireless_comm}

\begin{figure}[!ht]
    \subfloat[Average]{
    \includegraphics[width=0.23\textwidth]{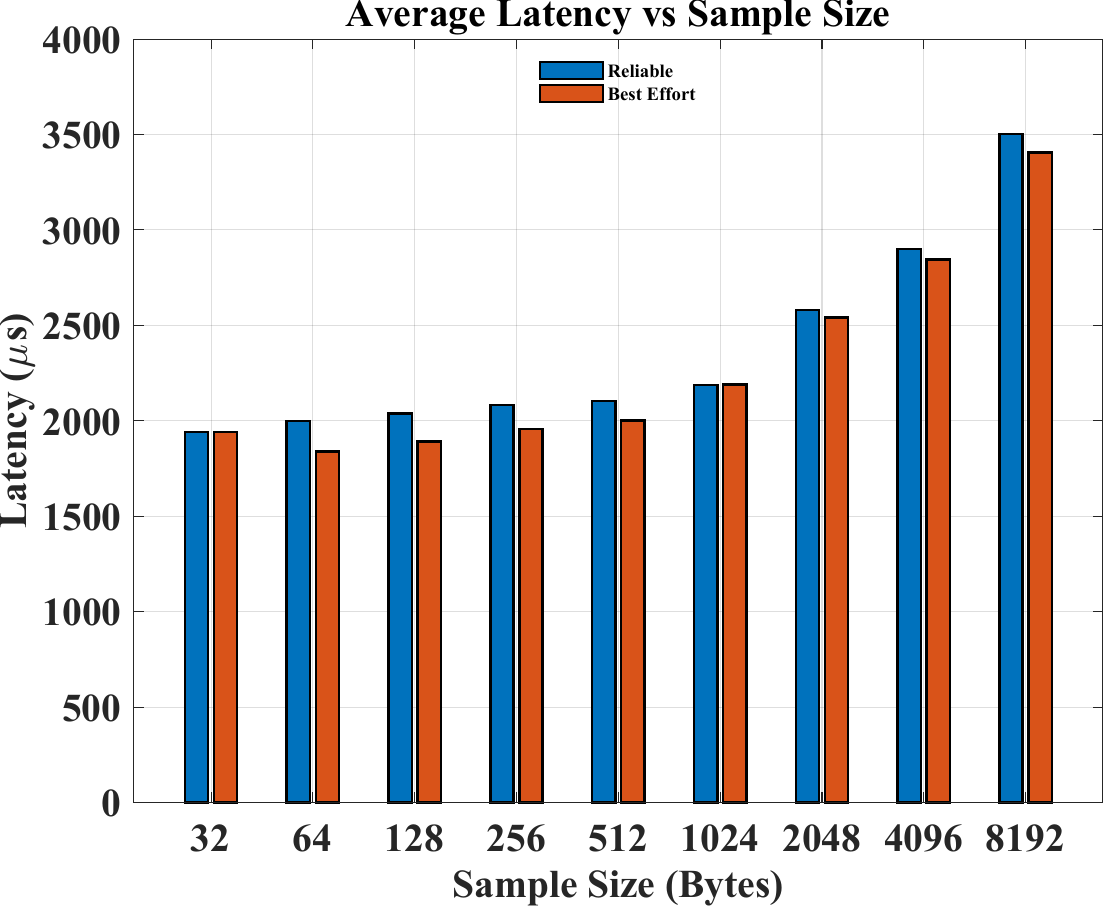}
    \label{fig:avg_latency_wireless}
    }\hfill
    \subfloat[Minimum]{
    \includegraphics[width=0.23\textwidth]{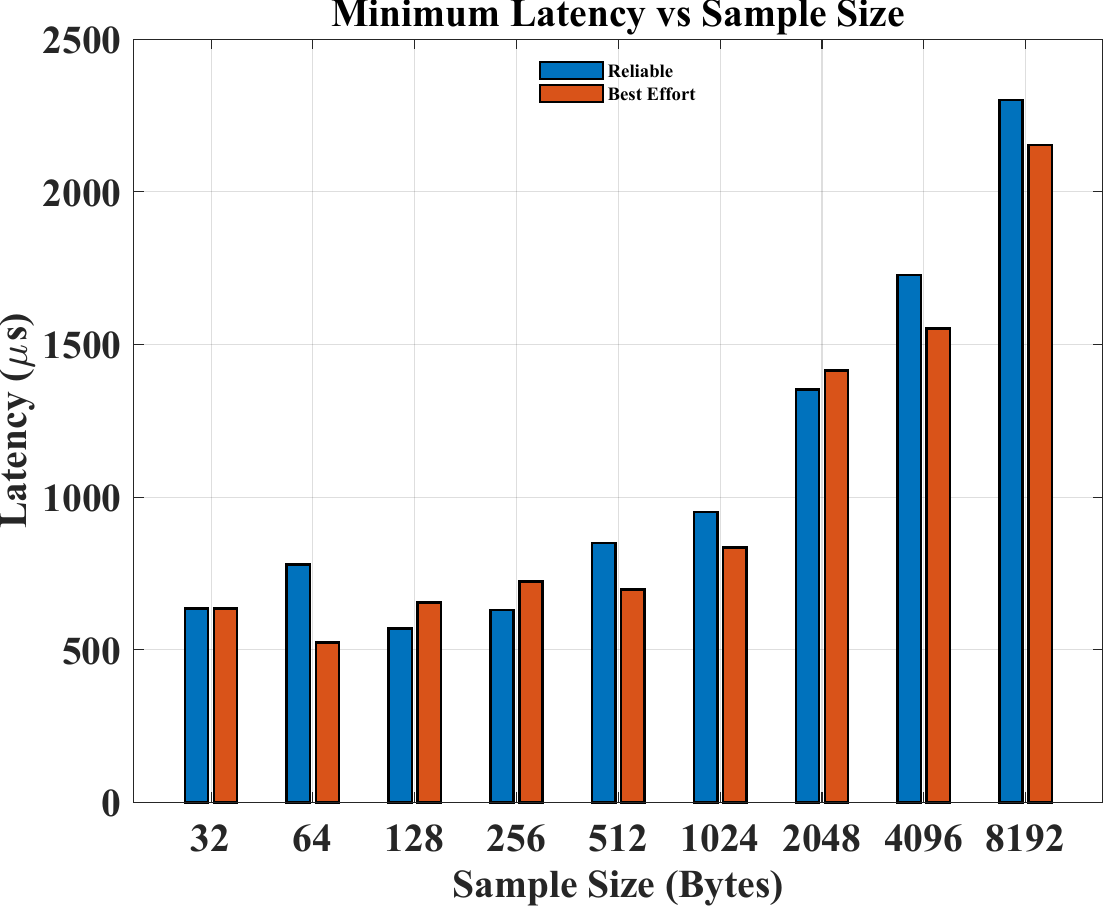}
    \label{fig:min_latency_wireless}
    }\\
    \subfloat[Maximum]{
    \includegraphics[width=0.23\textwidth]{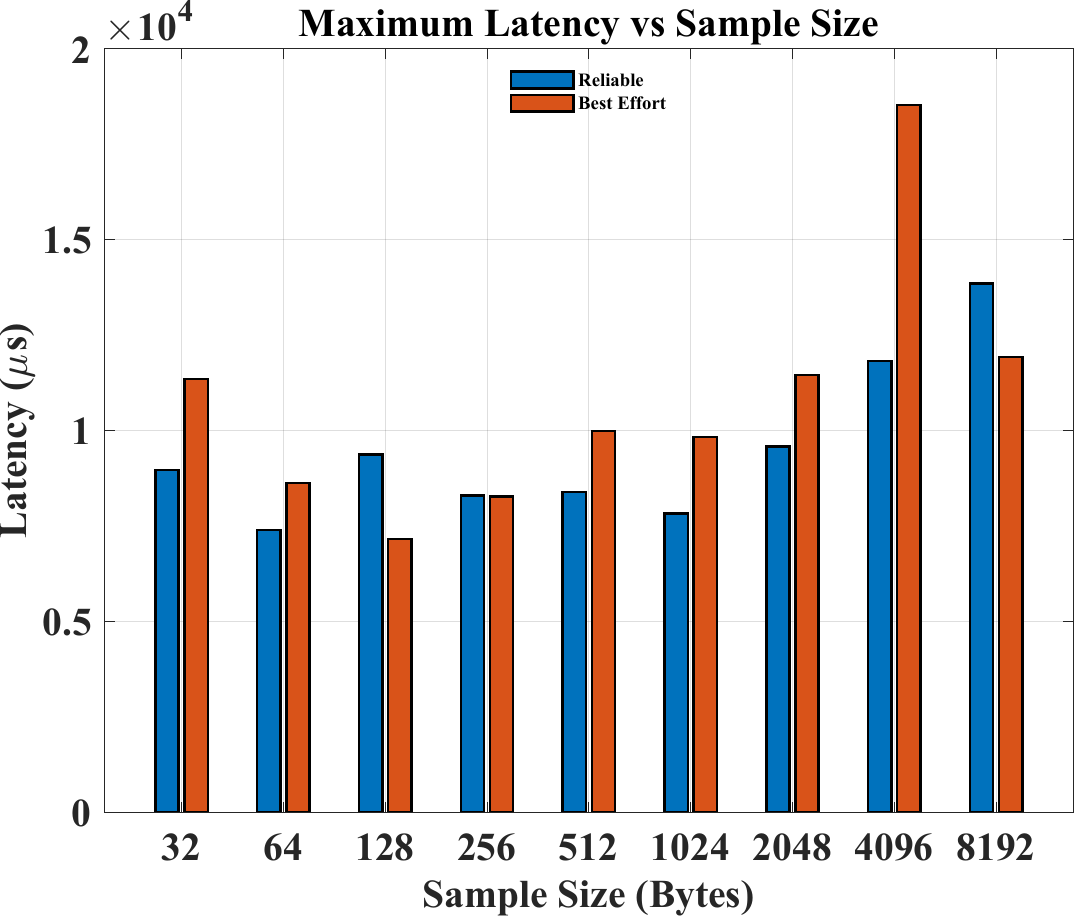}
    \label{fig:max_latency_wireless}
    }\hfill
    \subfloat[Standard Deviation]{
    \includegraphics[width=0.23\textwidth]{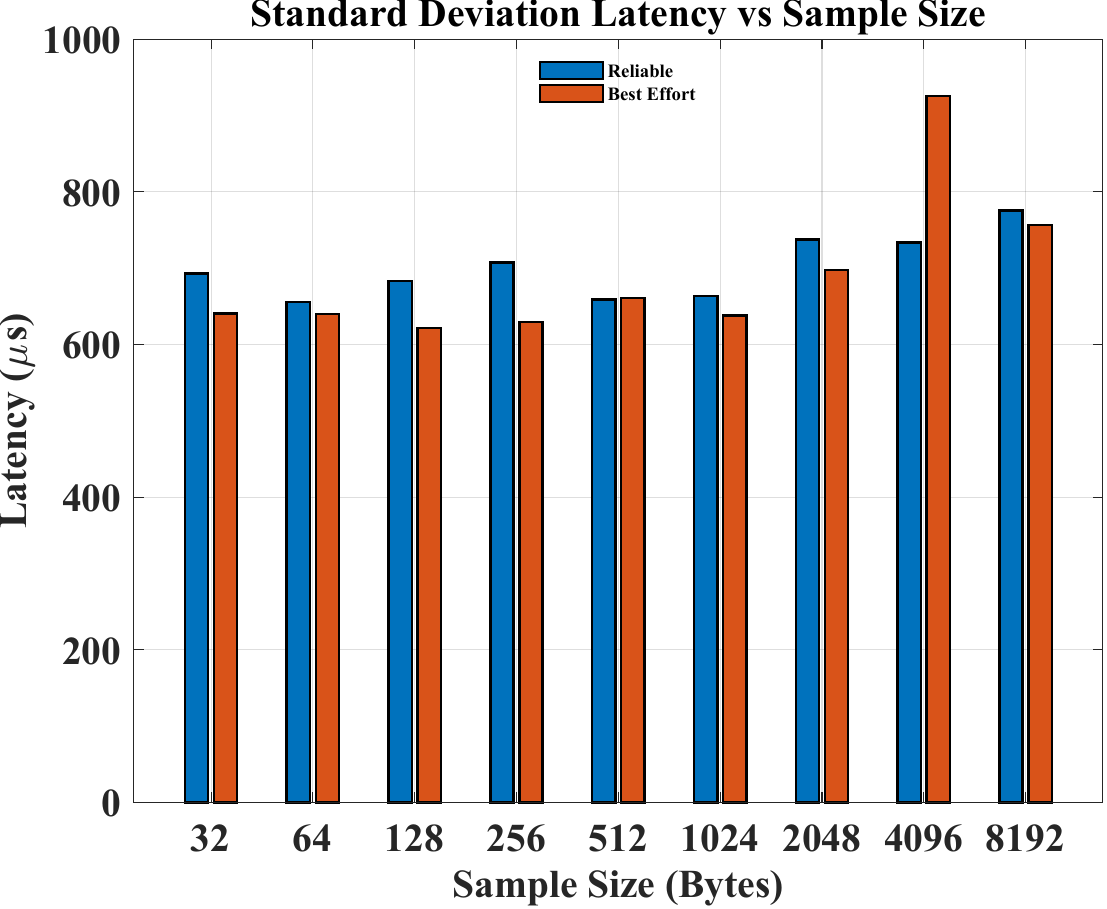}
    \label{fig:std_latency_wireless}
    }
    \caption{Latency of Proposed System using DDS over a Wireless Medium}
    \label{fig:latency-comparison-wireless}
\end{figure}

\begin{figure}[!ht]
    \subfloat[Total Samples]{
    \includegraphics[width=0.23\textwidth]{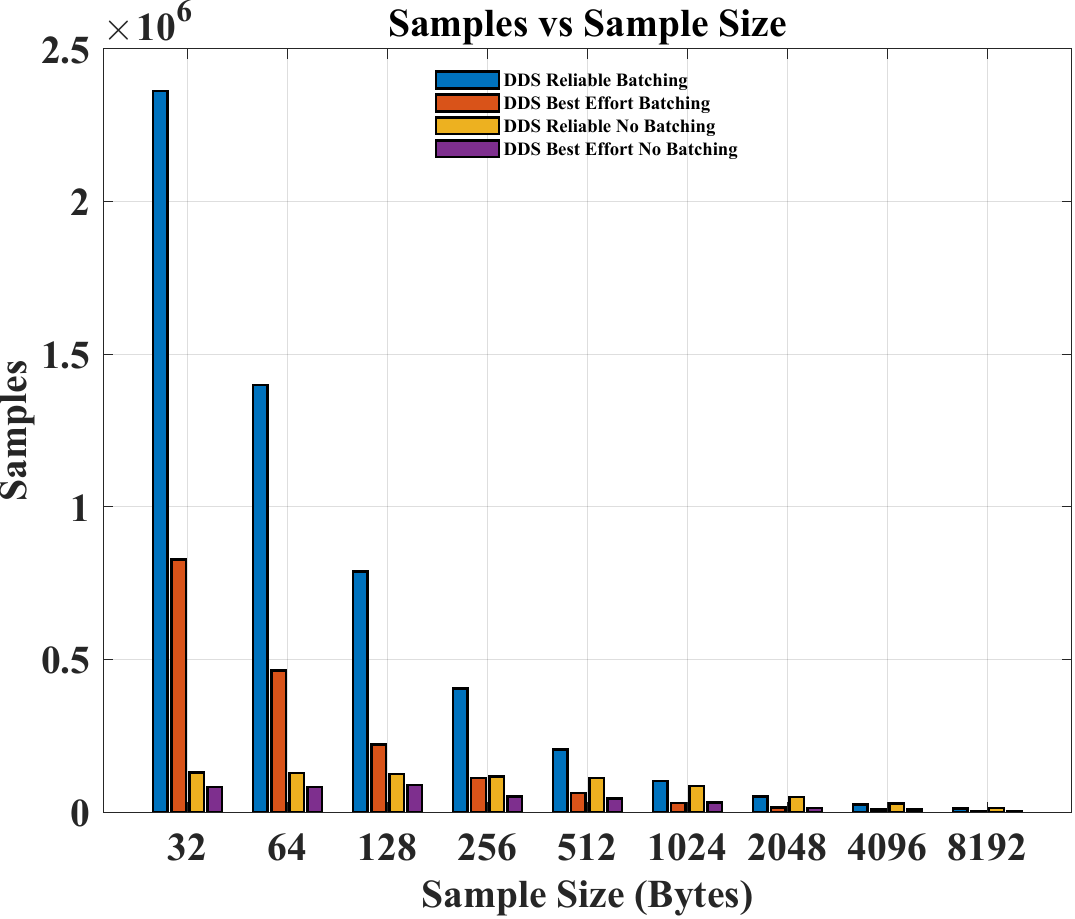}
    \label{fig:total_samples_wireless}
    }\hfill
    \subfloat[Average Samples]{
    \includegraphics[width=0.23\textwidth]{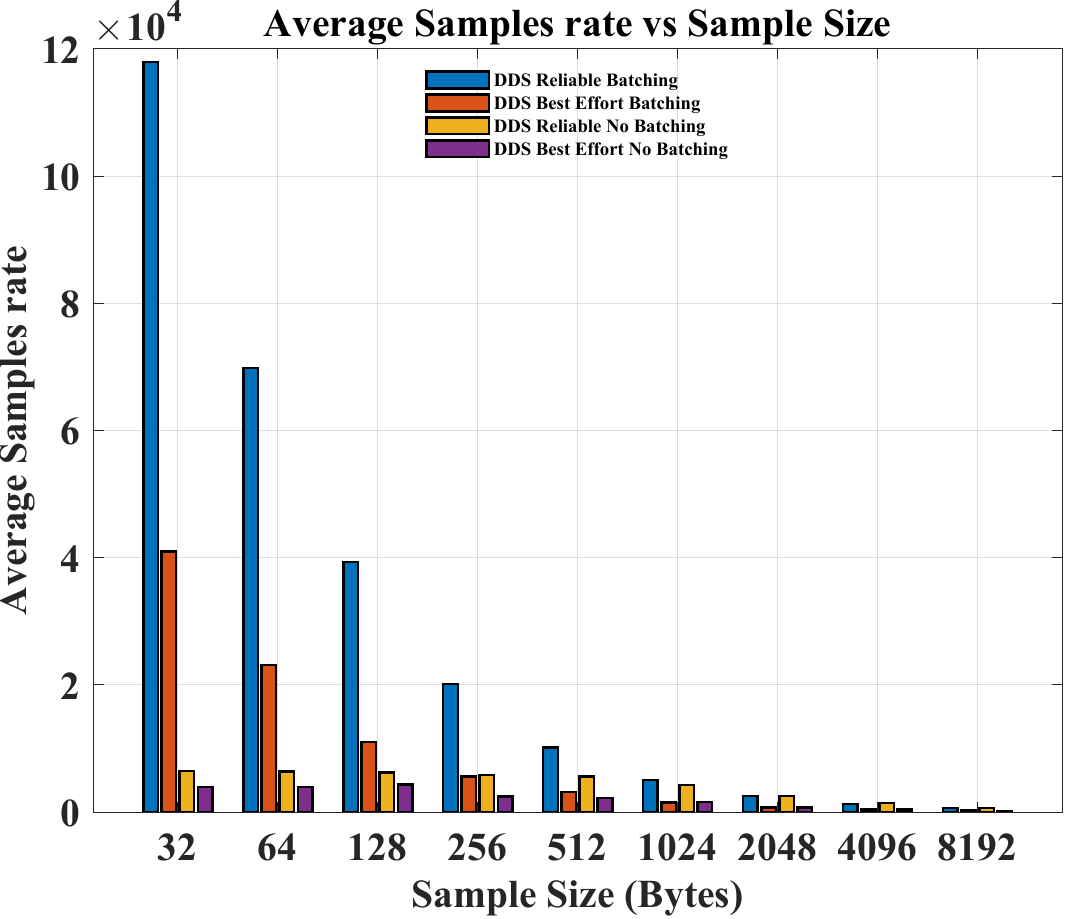}
    \label{fig:ave_sample_wireless}
    }\\
    \subfloat[Lost Samples]{
    \includegraphics[width=0.23\textwidth]{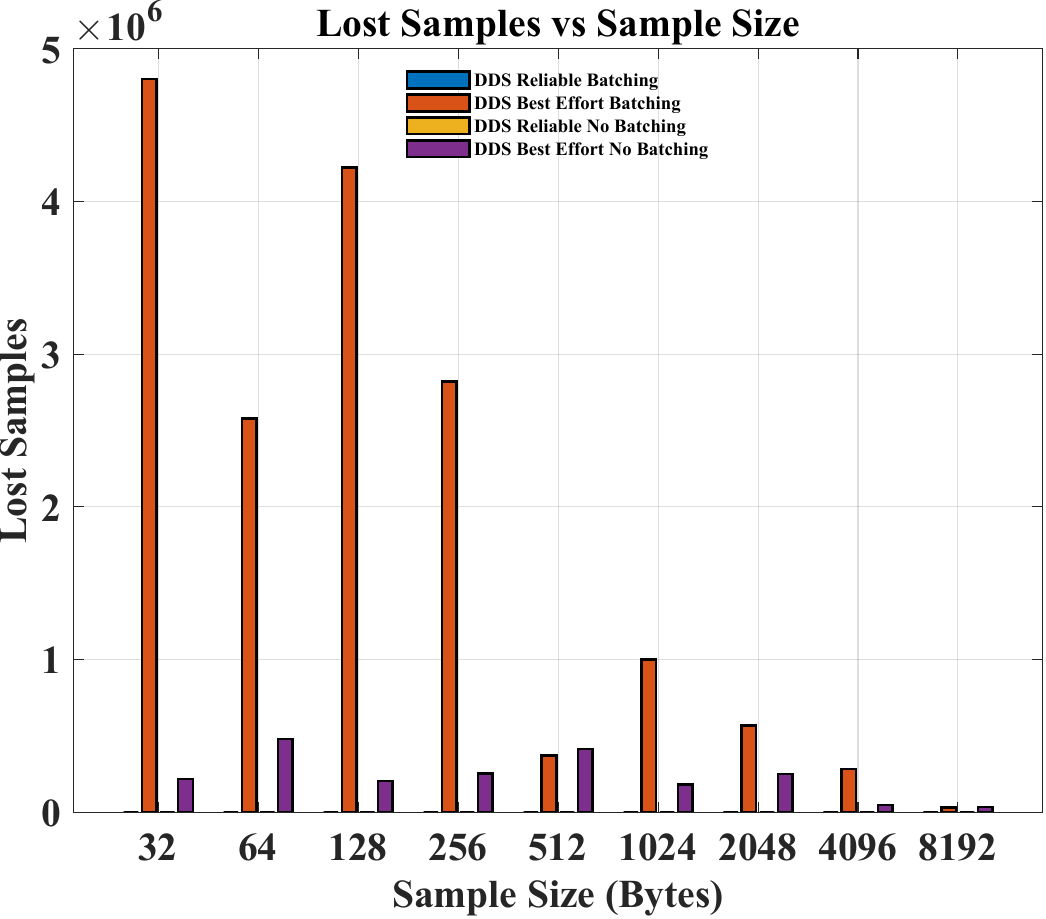}
    \label{fig:lost_samples_wireless}
    }\hfill
    \subfloat[Average Throughput]{
    \includegraphics[width=0.23\textwidth]{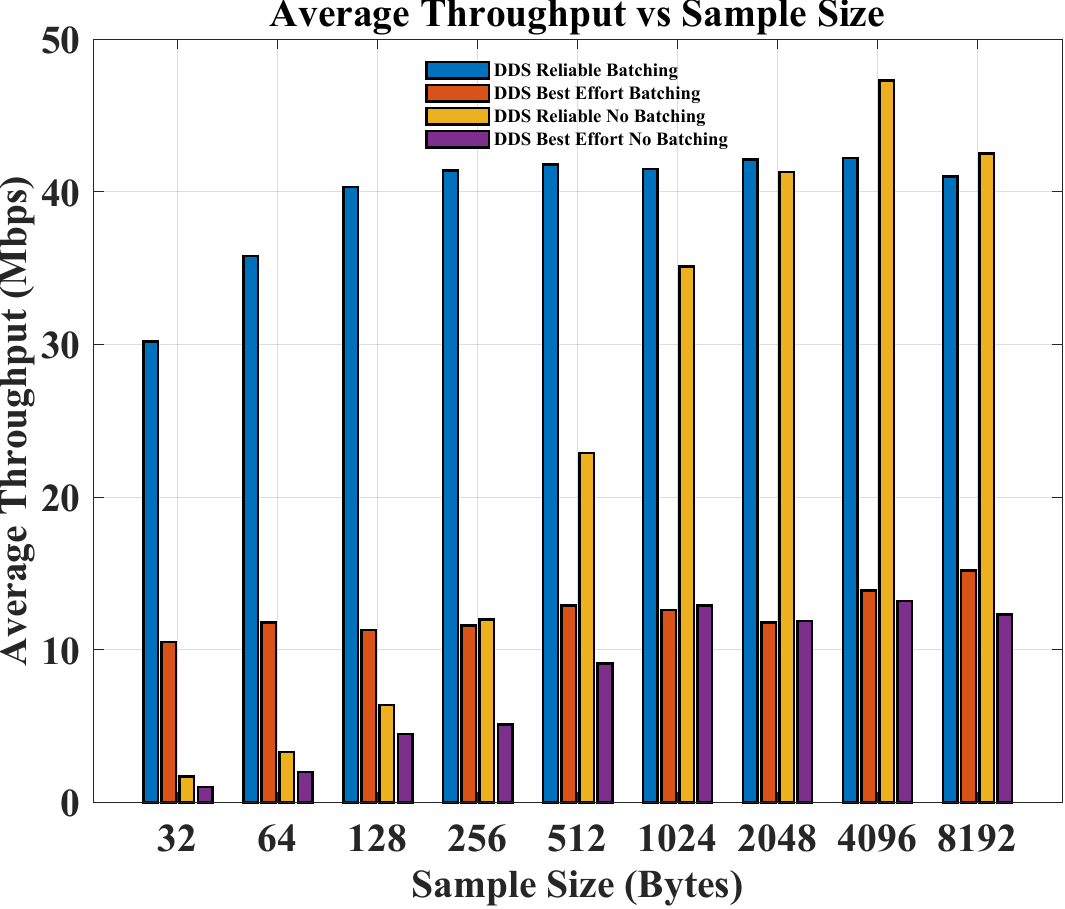}
    \label{fig:avg_throughput_wireless}
    }
    \caption{Other performance metrics of the proposed system using DDS over a Wireless Medium}
    \label{fig:throughput-comparison-wireless}
\end{figure}

The performance of the system over a wireless link is illustrated in Figs.~\ref{fig:latency-comparison-wireless} and \ref{fig:throughput-comparison-wireless}. The experimental setup consists of a TP-Link wireless router with a nominal data rate of 150~Mbps and two desktop computers equipped with 150~Mbps USB wireless adapters operating in IEEE 802.11b/g/n mode.

Average latency results for Reliable and Best Effort QoS are shown in Fig.~\ref{fig:avg_latency_wireless}. For small payloads up to 512~bytes, the average latency ranges between approximately 1.9 and 2.1~ms. As payload size increases, latency gradually rises and reaches about 3.4 to 3.5~ms at 8192~bytes. This trend follows the increased transmission time of larger frames and the fragmentation that occurs when payloads exceed the 1500~byte MTU. Minimum latency values in Fig.~\ref{fig:min_latency_wireless} represent best-case channel conditions with limited contention, while maximum latency and standard deviation in Figs.~\ref{fig:max_latency_wireless} and \ref{fig:std_latency_wireless} reflect variability introduced by wireless channel access and operating system scheduling. Reliable QoS incurs additional processing for sample acknowledgment and recovery, which slightly increases latency variation compared to Best Effort.

The throughput-related metrics are presented in Fig.~\ref{fig:throughput-comparison-wireless}. The total number of delivered samples is shown in Fig.~\ref{fig:total_samples_wireless}. When batching is enabled under Reliable QoS, the number of successfully delivered samples increases substantially, particularly for small payloads. For example, at 32~bytes, batching allows the delivery of more than two million samples, compared to a much smaller count when batching is disabled. Average samples per second, shown in Fig.~\ref{fig:ave_sample_wireless}, follow the same trend. Average throughput in Fig.~\ref{fig:avg_throughput_wireless} remains between roughly 30 and 42~Mbps for Reliable QoS with batching across all payload sizes. Without batching, throughput increases with payload size up to 4096~bytes and then decreases slightly at 8192~bytes due to fragmentation and reassembly overhead.

Packet loss over the wireless link is shown in Fig.~\ref{fig:lost_samples_wireless}. Reliable QoS maintains zero packet loss across all tested payload sizes and batching configurations. In contrast, Best Effort QoS exhibits high loss rates, particularly for small payloads, where loss exceeds 70~percent and approaches 90~percent at larger payload sizes. This behavior is consistent with the lack of retransmission or recovery mechanisms in Best Effort mode. While batching increases the number of transmitted samples, it does not reduce the observed loss ratios for Best Effort, and the number of samples received by the subscriber remains limited.  Unlike in the wired network, where our system almost utilizes the entire available
bandwidth, with throughput reaching up to approximately 0.94 Gbps for Reliable and 0.94 Gbps
for Best Effort QoS as shown in Fig. 6(d), the throughput in the wireless network does not converge
to the physical bandwidth of 150 Mbps as the payload increases, which we believe is due to the
limitations of the USB Wi Fi connector and our TP Link router.

\subsection{Impact of DDS Security Overhead}
\label{subsec:dds-sec}

The impact of enabling DDS Security was evaluated on both wired Ethernet and wireless links. The security configuration applies confidentiality and integrity protection to RTPS messages by encrypting each data sample and attaching an integrity tag.

\begin{figure}[!ht]
    \subfloat[Average]{
    \includegraphics[width=0.23\textwidth]{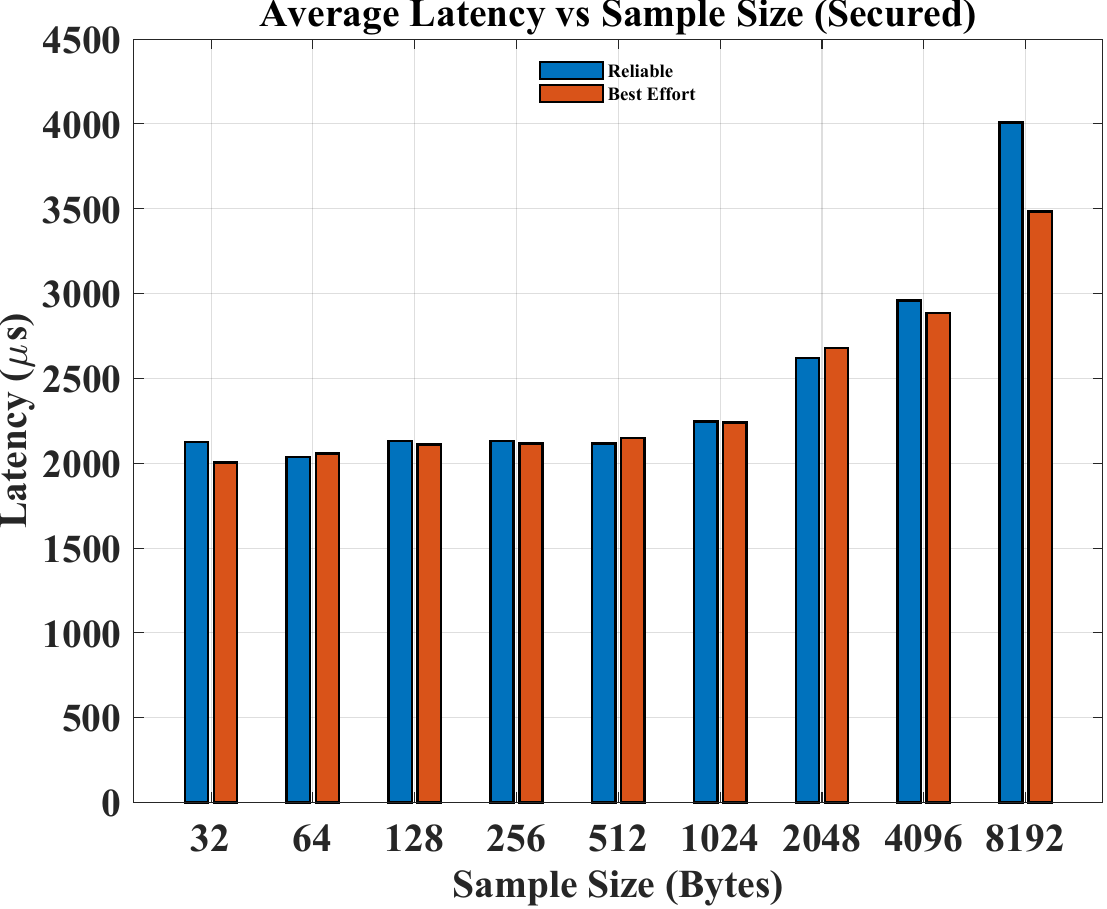}
    \label{fig:avg_latency_wireless_sec}
    }\hfill
    \subfloat[Minimum]{
    \includegraphics[width=0.23\textwidth]{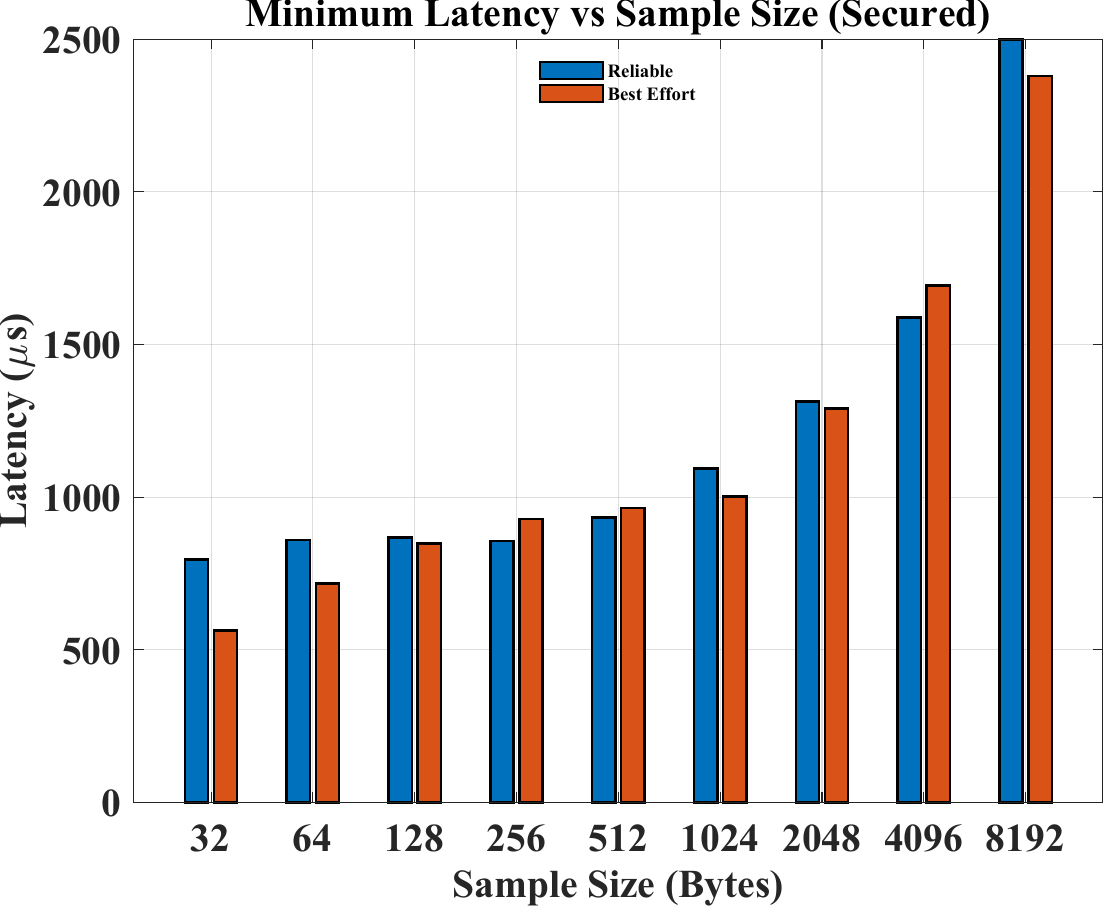}
    \label{fig:min_latency_wireless_sec}
    }\\
    \subfloat[Maximum]{
    \includegraphics[width=0.23\textwidth]{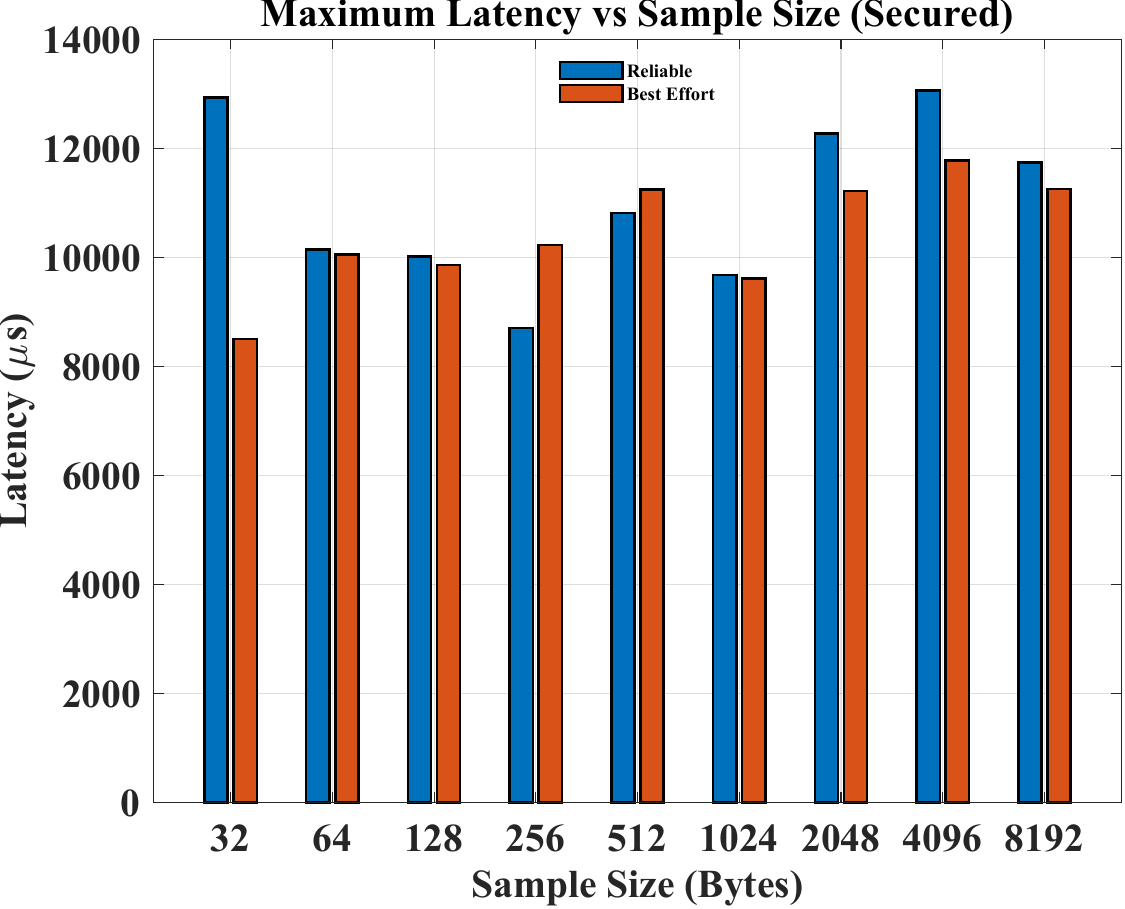}
    \label{fig:max_latency_wireless_sec}
    }\hfill
    \subfloat[Standard Deviation]{
    \includegraphics[width=0.23\textwidth]{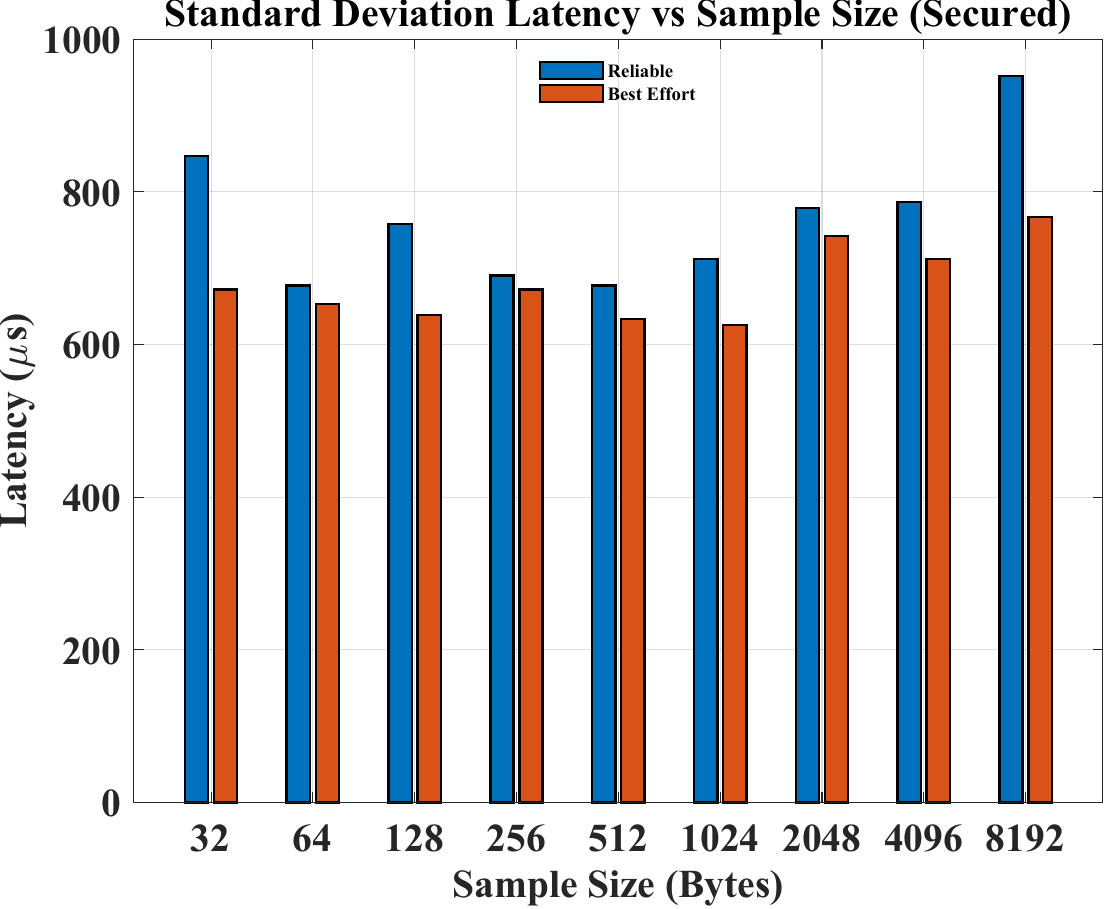}
    \label{fig:std_latency_wireless_sec}
    }
    \caption{Latency of Proposed System using DDS over a Secured Wireless Medium}
    \label{fig:latency-comparison-wireless-sec}
\end{figure}

Latency results for the secured wireless link are presented in Fig.~\ref{fig:latency-comparison-wireless-sec}. Average latency values in Fig.~\ref{fig:avg_latency_wireless_sec} increase slightly for both Reliable and Best Effort QoS compared to the unsecured case. For example, at 32~bytes, Reliable QoS latency increases from approximately 1.9 to 3.5~ms at 8192~bytes to about 2.1 to 4.0~ms. Minimum, maximum, and standard deviation values in Figs.~\ref{fig:min_latency_wireless_sec}, \ref{fig:max_latency_wireless_sec}, and \ref{fig:std_latency_wireless_sec} remain close to those observed without security. The variability introduced by the wireless medium dominates the overall delay, reducing the relative impact of cryptographic processing.

\begin{figure}[!ht]
    \subfloat[Average]{
    \includegraphics[width=0.23\textwidth]{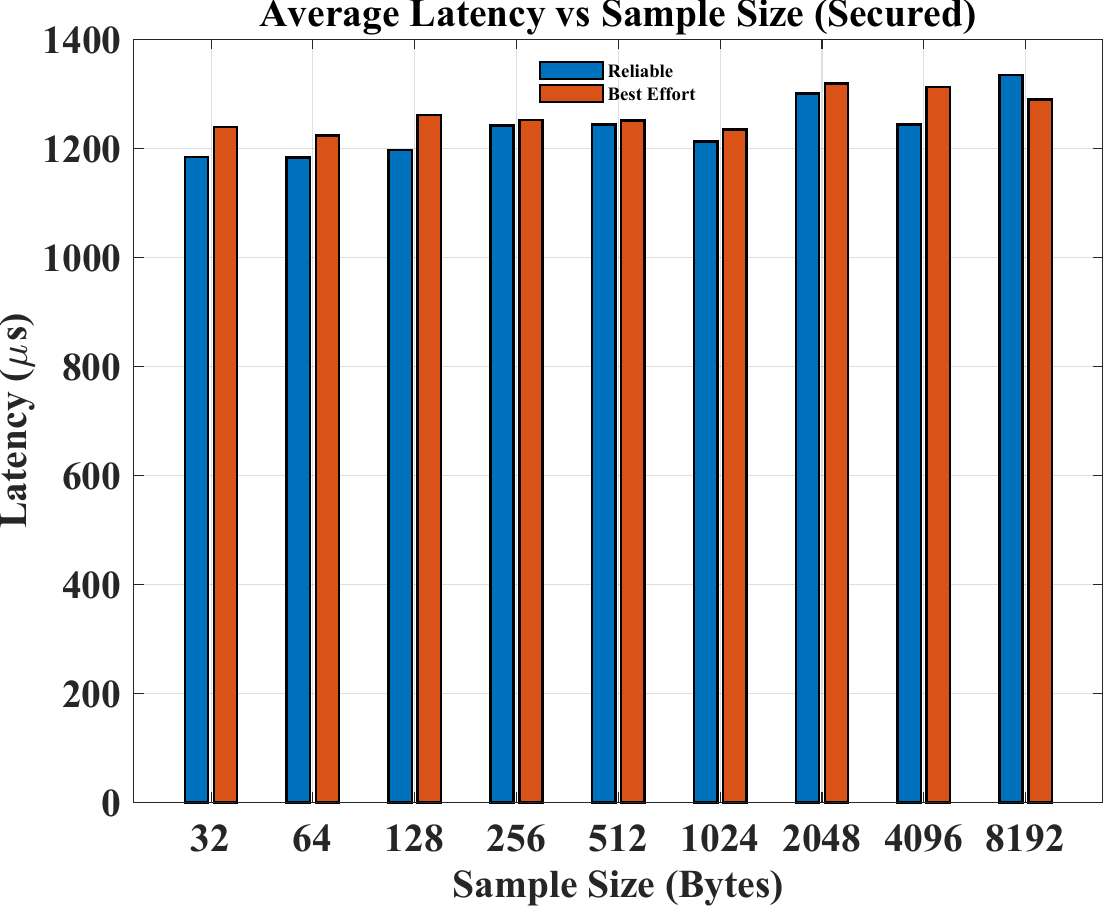}
    \label{fig:avg_latency_wired_sec}
    }\hfill
    \subfloat[Minimum]{
    \includegraphics[width=0.23\textwidth]{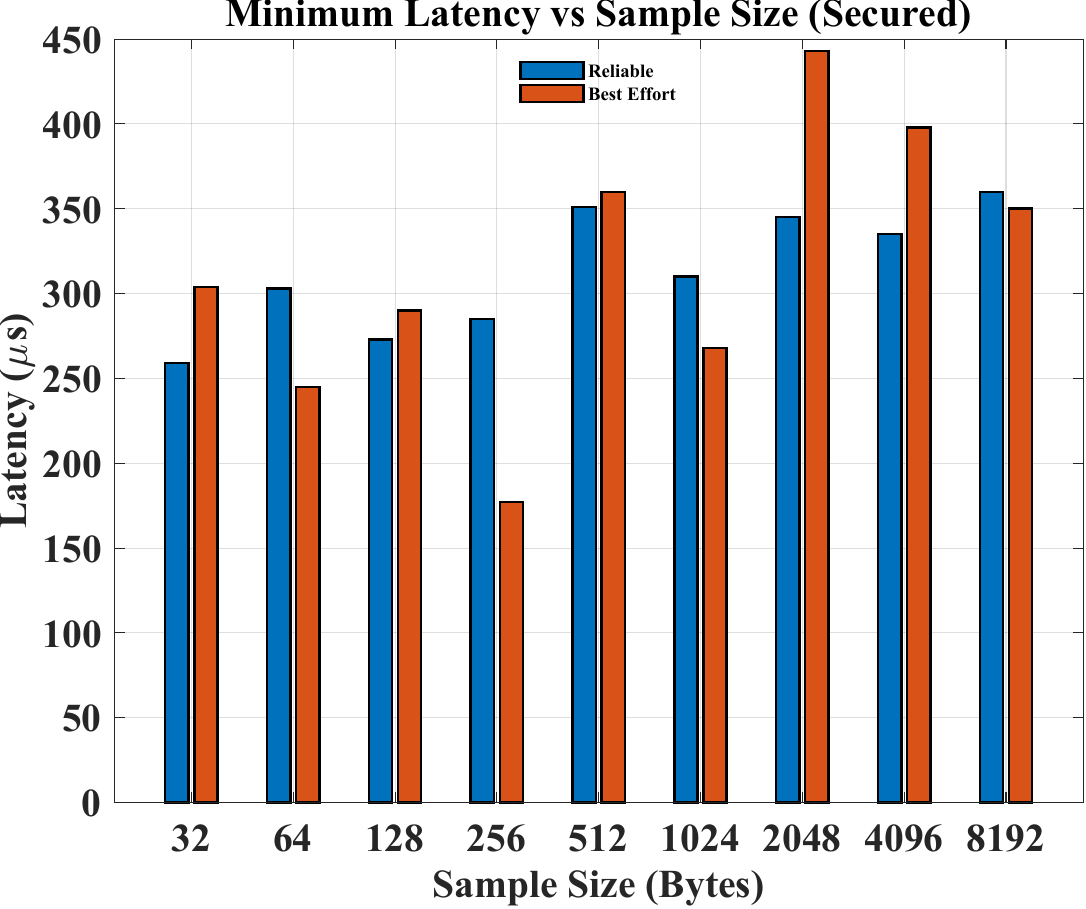}
    \label{fig:min_latency_wired_sec}
    }\\
    \subfloat[Maximum]{
    \includegraphics[width=0.23\textwidth]{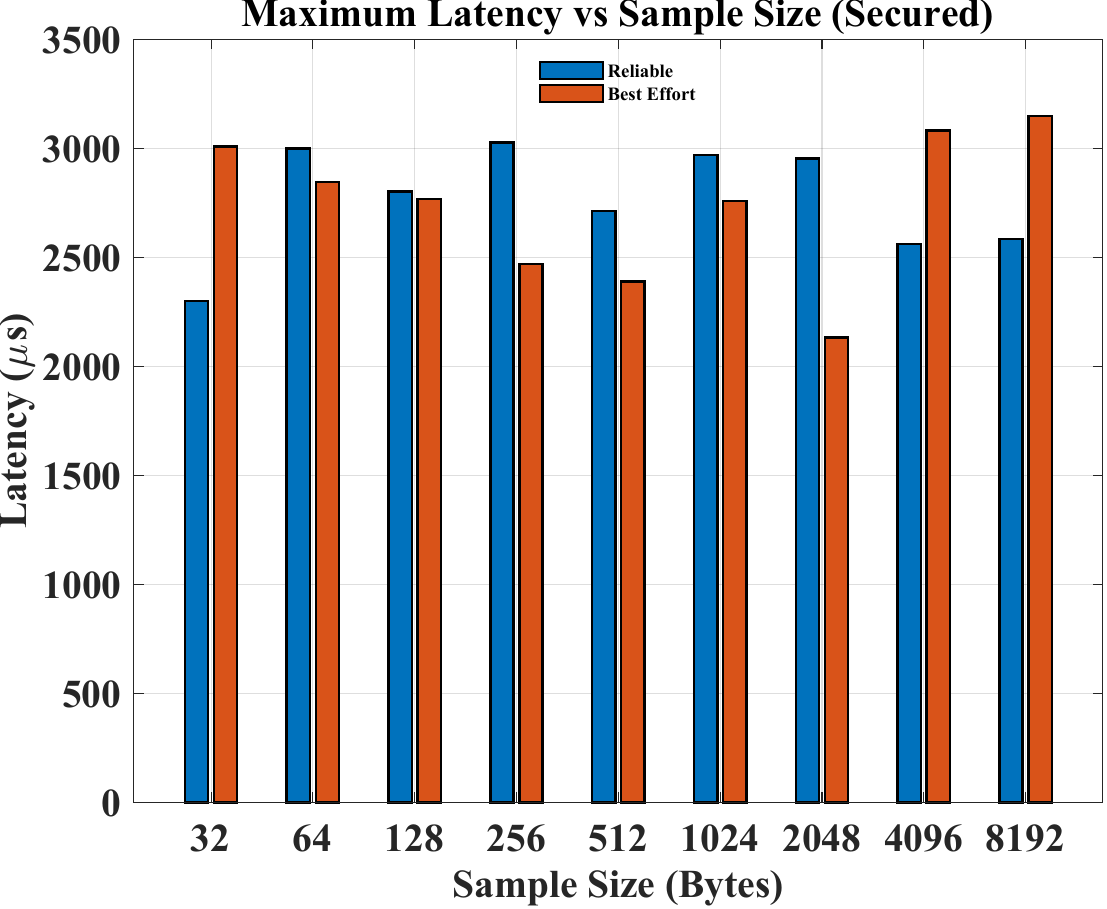}
    \label{fig:max_latency_wired_sec}
    }\hfill
    \subfloat[Standard Deviation]{
    \includegraphics[width=0.23\textwidth]{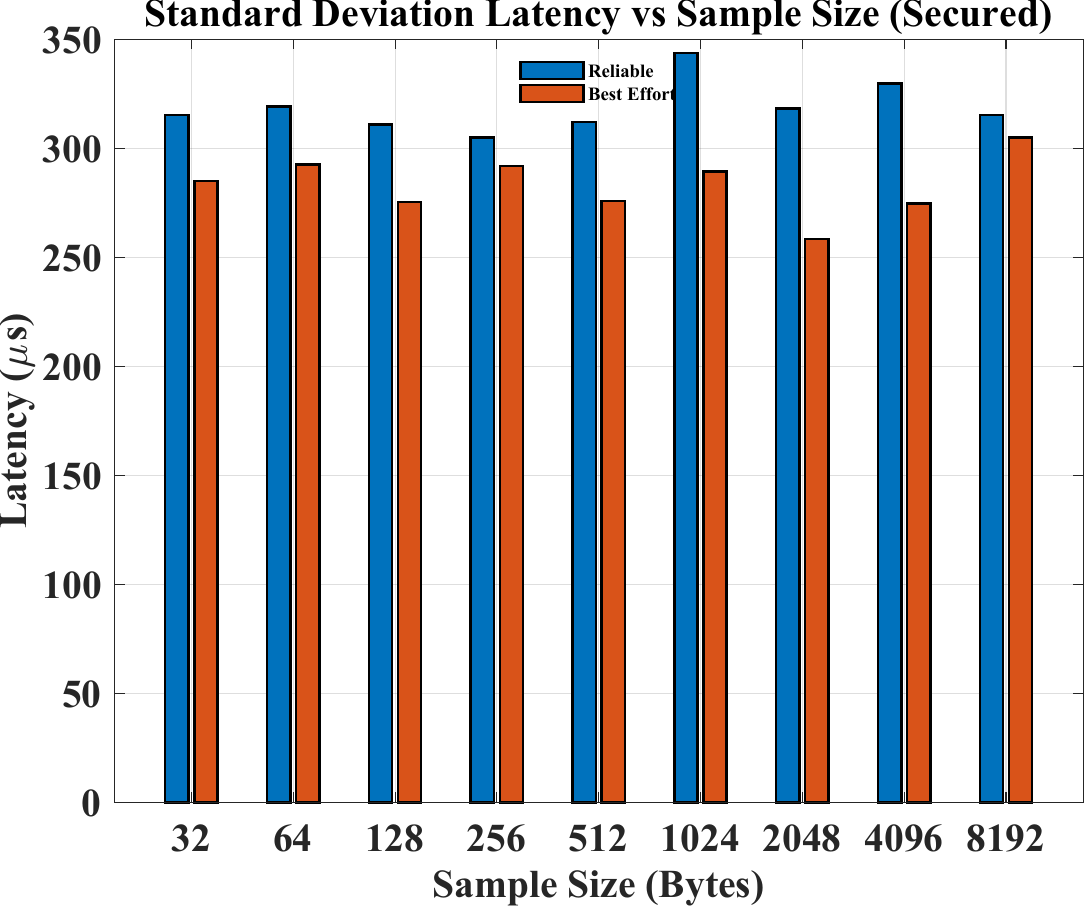}
    \label{fig:std_latency_wired_sec}
    }
    \caption{Latency of Proposed System using DDS over a Secured Ethernet Network}
    \label{fig:latency-comparison-wired-sec}
\end{figure}

Latency over a secured Ethernet link is shown in Fig.~\ref{fig:latency-comparison-wired-sec}. For small payloads, enabling security increases average latency due to the fixed cost of encryption and integrity checks applied per message. This effect is visible in Fig.~\ref{fig:avg_latency_wired_sec}, where latency increases notably at 32~bytes. As payload size increases, the difference between secured and unsecured latency decreases, since transmission time becomes the dominant factor. Minimum, maximum, and standard deviation results in Figs.~\ref{fig:min_latency_wired_sec}, \ref{fig:max_latency_wired_sec}, and \ref{fig:std_latency_wired_sec} show similar variability with and without security, reflecting operating system scheduling effects rather than cryptographic overhead.

\begin{figure}[!ht]
    \subfloat[Total Samples]{
    \includegraphics[width=0.23\textwidth]{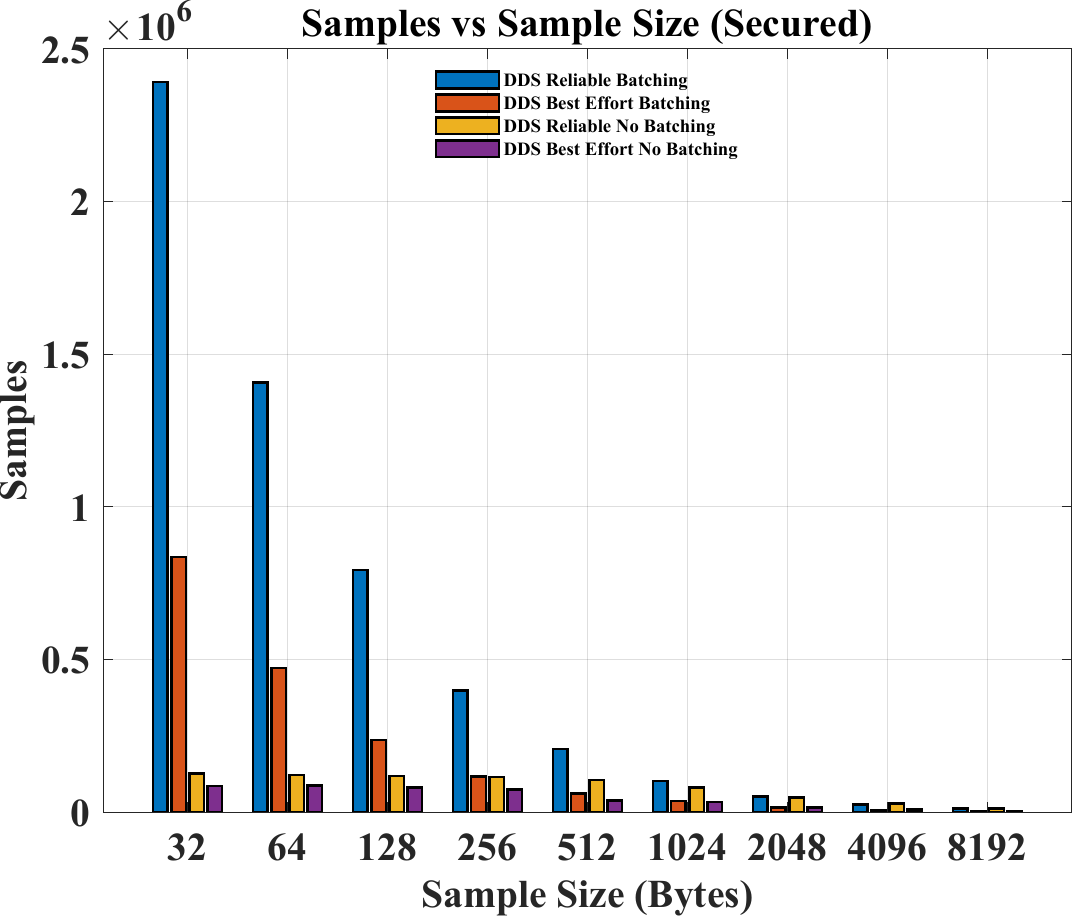}
    \label{fig:total_samples_wireless_sec}
    }\hfill
    \subfloat[Average Samples]{
    \includegraphics[width=0.23\textwidth]{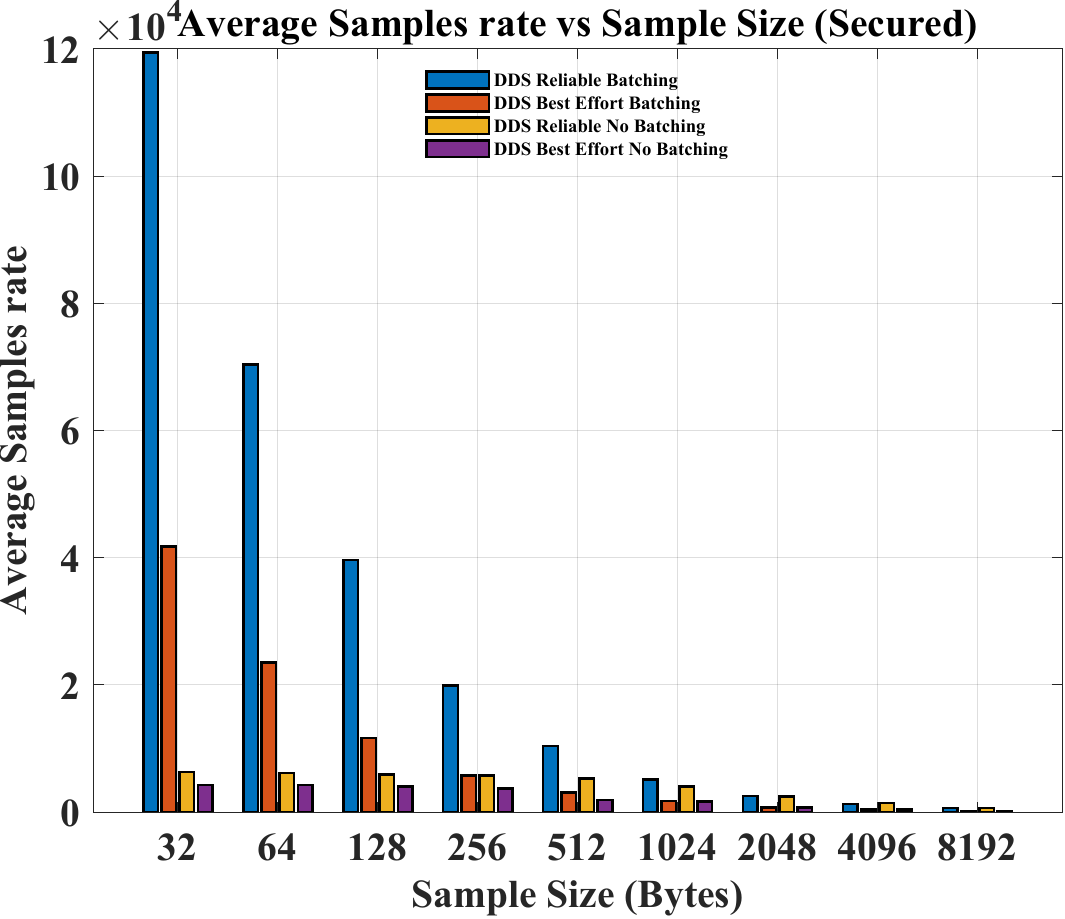}
    \label{fig:ave_sample_wireless_sec}
    }\\
    \subfloat[Lost Samples]{
    \includegraphics[width=0.23\textwidth]{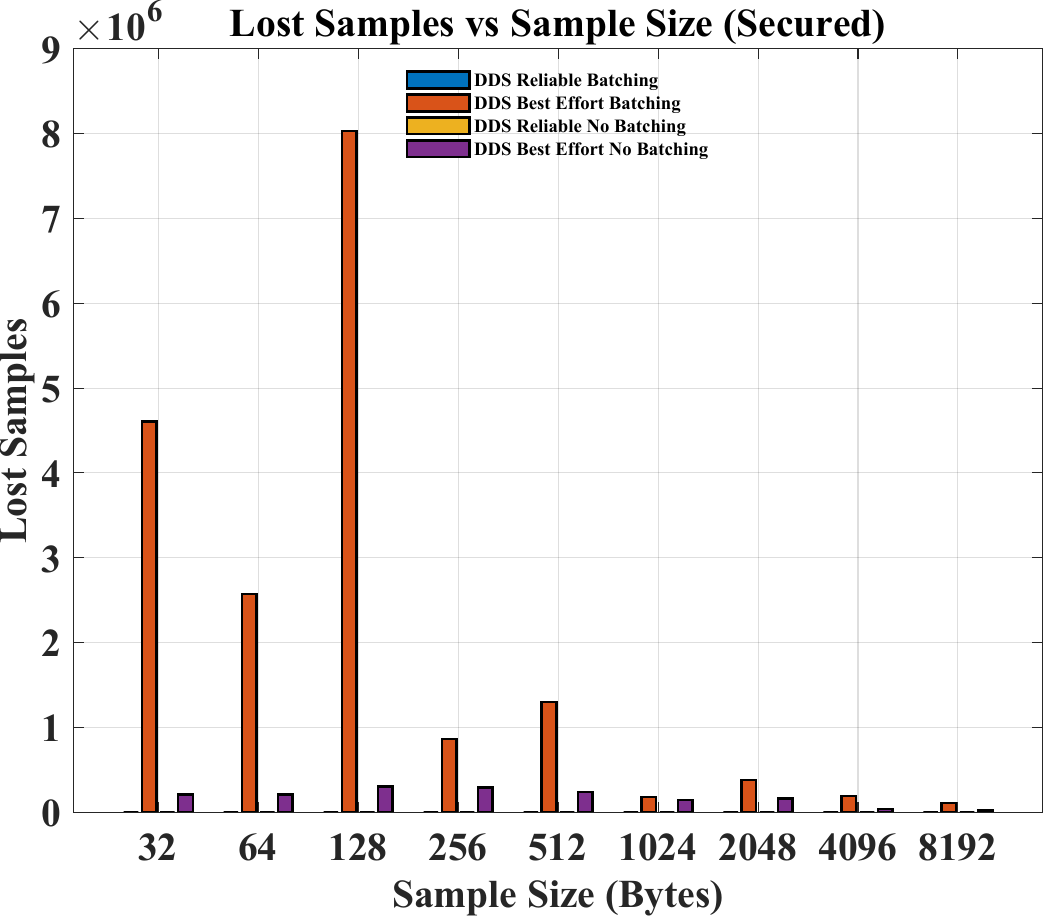}
    \label{fig:lost_samples_wireless_sec}
    }\hfill
    \subfloat[Average Throughput]{
    \includegraphics[width=0.23\textwidth]{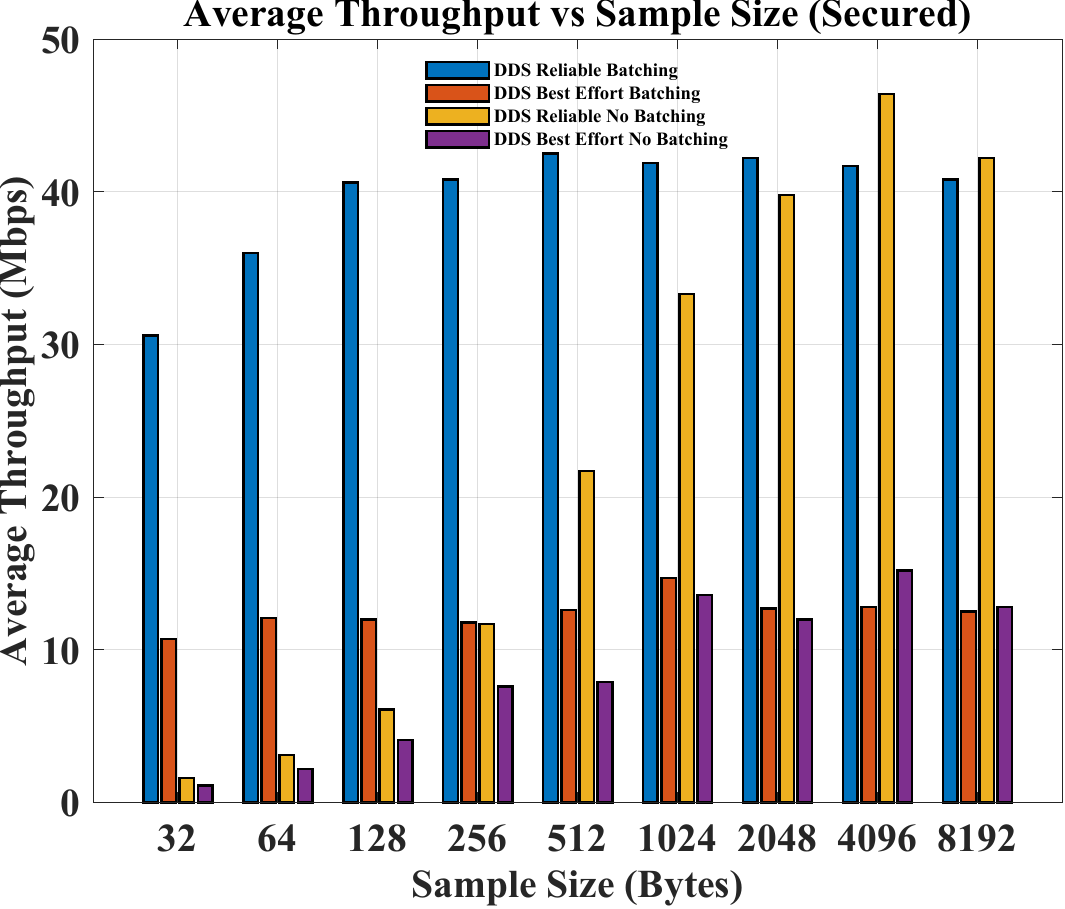}
    \label{fig:avg_throughput_wireless_sec}
    }
    \caption{Other performance metrics of the proposed system using DDS over a Secured Wireless Medium}
    \label{fig:throughput-comparison-wireless-sec}
\end{figure}

Throughput and loss metrics for the secured wireless link are shown in Fig.~\ref{fig:throughput-comparison-wireless-sec}. The total number of delivered samples and average samples per second, shown in Figs.~\ref{fig:total_samples_wireless_sec} and \ref{fig:ave_sample_wireless_sec}, closely match the unsecured wireless results. Reliable QoS maintains zero packet loss, as shown in Fig.~\ref{fig:lost_samples_wireless_sec}, while Best Effort continues to experience high loss rates. Average throughput in Fig.~\ref{fig:avg_throughput_wireless_sec} remains between approximately 30 and 41~Mbps for Reliable QoS with batching, indicating that the impact of security is limited when batching is used. This is because batching allows
cryptographic protection to be applied to an entire batch, reducing the number of cryptographic
operations and thus the overhead per message.

\begin{figure}[!ht]
    \subfloat[Total Samples]{
    \includegraphics[width=0.23\textwidth]{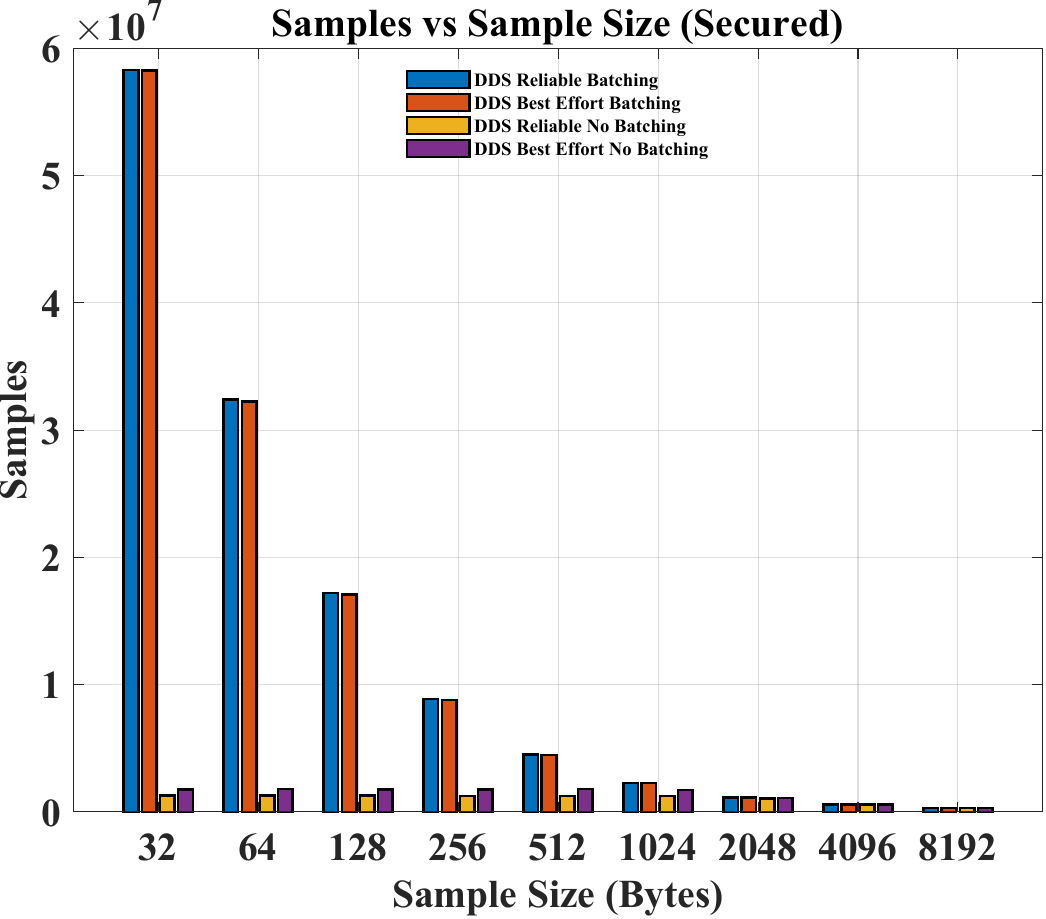}
    \label{fig:total_samples_wired_sec}
    }\hfill
    \subfloat[Average Samples]{
    \includegraphics[width=0.23\textwidth]{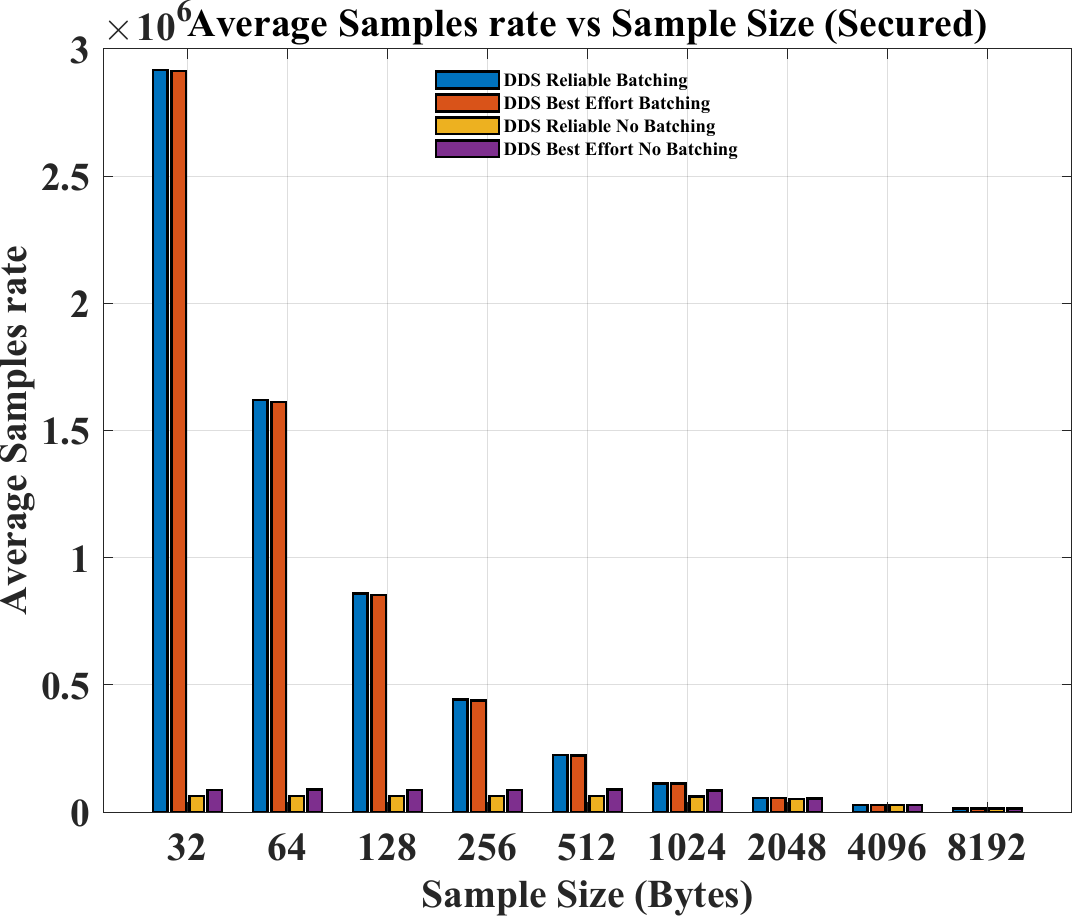}
    \label{fig:ave_sample_wired_sec}
    }\\
    \subfloat[Average Throughput]{
    \includegraphics[width=0.23\textwidth]{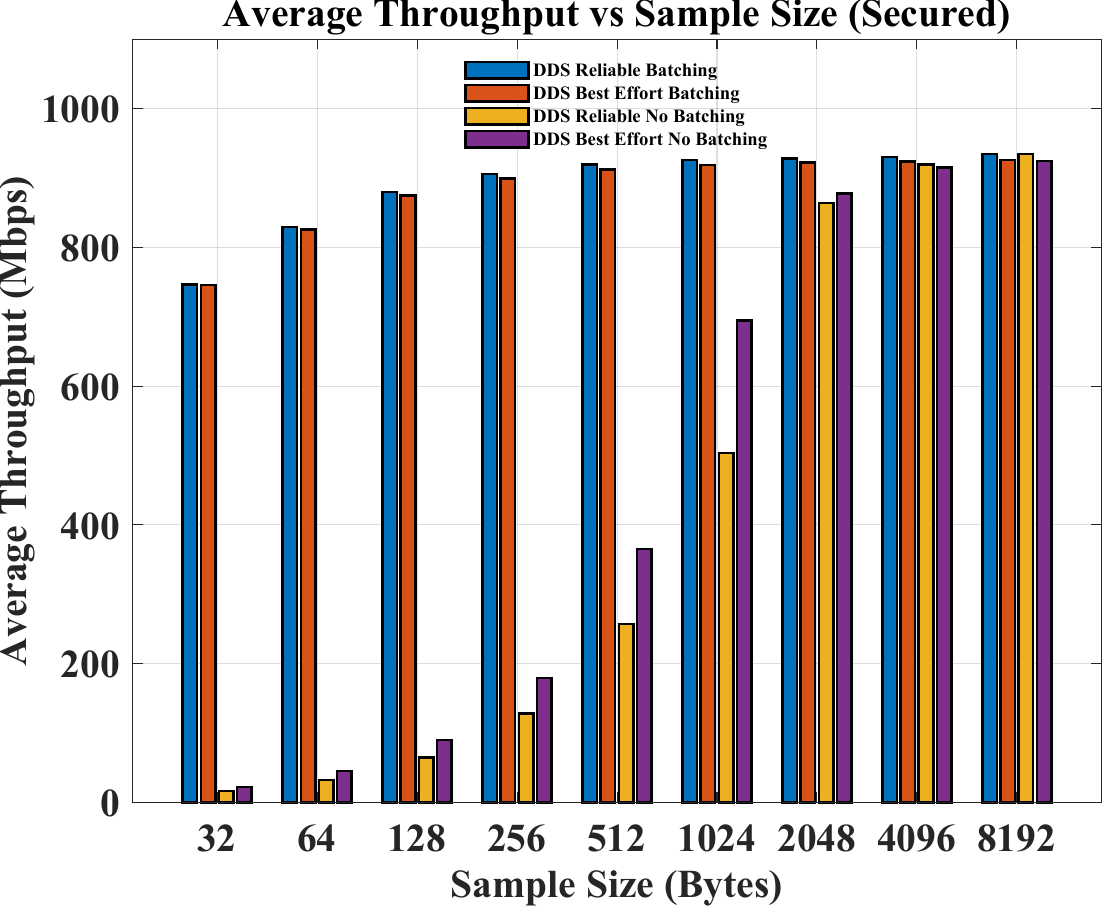}
    \label{fig:avg_throughput_wired_sec}
    }
    \caption{Other performance metrics of the proposed system using DDS over a Secured Ethernet}
    \label{fig:throughput-comparison-wired-sec}
\end{figure}

For the secured Ethernet link, throughput metrics are shown in Fig.~\ref{fig:throughput-comparison-wired-sec}. The total number of samples and average throughput in Figs.~\ref{fig:total_samples_wired_sec}, \ref{fig:ave_sample_wired_sec}, and \ref{fig:avg_throughput_wired_sec} show that encryption reduces throughput for small payloads when batching is disabled, due to per-message processing overhead. When batching is enabled, throughput remains high and reaches several hundred megabits per second, with minimal difference between secured and unsecured modes.

\section{Conclusion}
\label{sec:conc}

This paper examined the feasibility of employing DDS middleware as the communication backbone for a heterogeneous real-time patient monitoring system. The proposed system architecture consists of modular DDS domain participants deployed across patient rooms and ward-level applications, connected through a layered databus structure. Performance metrics such as latency, throughput, and packet loss were measured across both wired and wireless LAN environments. Quality of Service (QoS) policies, including Reliable and Best Effort, were evaluated to examine the trade-offs between delivery guarantees and communication overhead. The results showed that Reliable QoS achieved lossless delivery in all experiments, while Best Effort provided slightly lower latency at the cost of occasional data loss. The evaluation further highlighted the benefits of batching in improving throughput, reinforcing the importance of QoS selection for different clinical scenarios. Overall, the findings confirm that DDS is a practical and effective middleware for real-time clinical communication, particularly in use cases where consistent and reliable data updates are essential. For future work, we plan to extend the experiments to assess system scalability and evaluate performance over wide-area networks (WANs) for large-scale healthcare deployments.

\section*{Acknowledgments}
The authors would like to acknowledge all support provided by Real-Time Innovations (RTI), Alfozan Academy, the Applied Research Center for Non-Proft and Social Development (ARCS), and King Fahd University of Petroleum \& Minerals (KFUPM).

 \bibliographystyle{IEEEtran}
\bibliography{refs}


\vspace{11pt}
\begin{IEEEbiography}[{\includegraphics[width=1in,height=1.25in,clip,keepaspectratio]{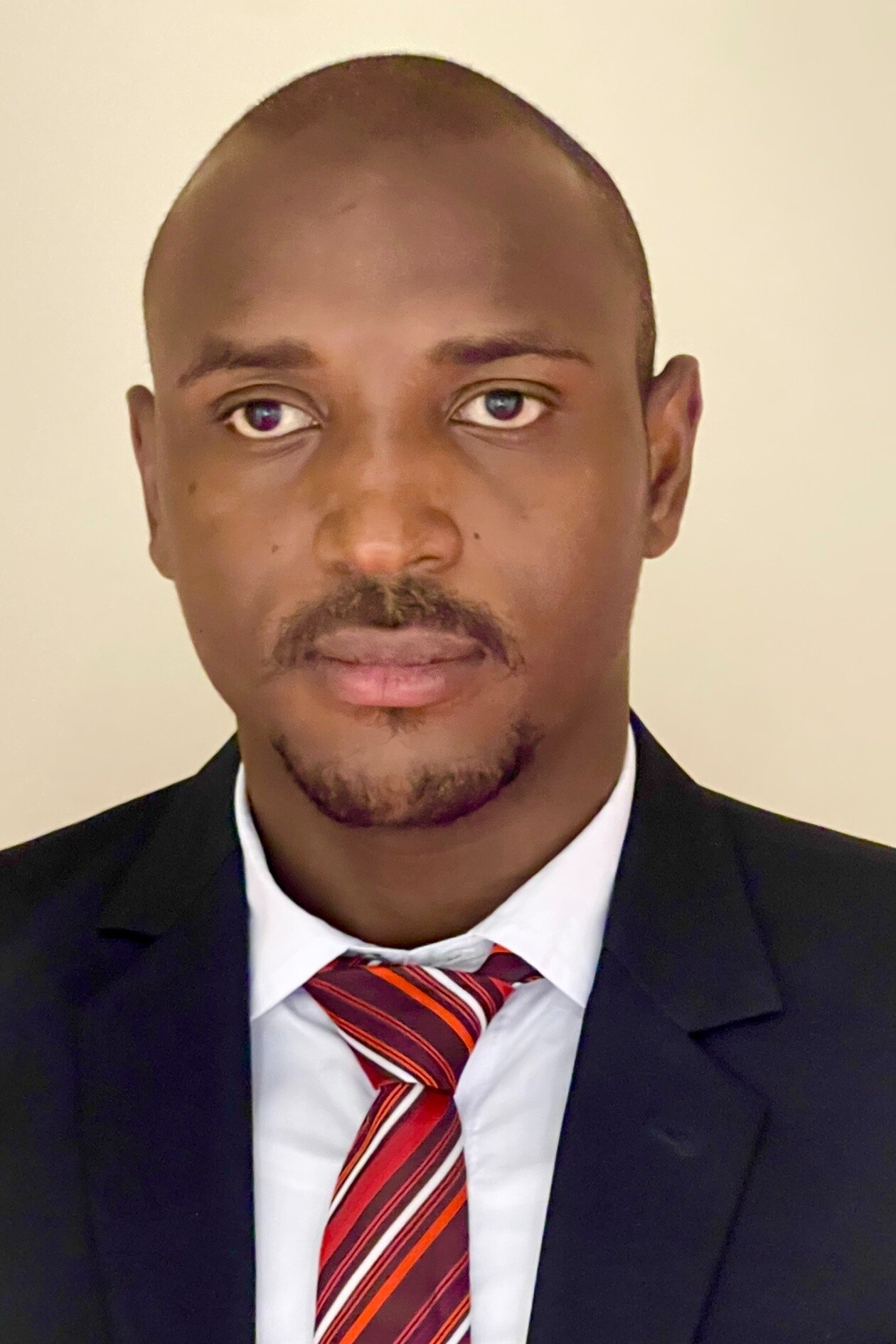}}]{Muhammad Dikko Gambo} received his B.ScTE degree in Computer Science and Engineering from the Islamic University of Technology, Dhaka, Bangladesh, and a National Innovation Diploma (NID) in Networking and System Security from the Katsina State Institute of Technology and Management, Katsina, Nigeria. He later completed the M.Sc. degree in Computer Engineering, with a concentration in Networks and Cybersecurity, at King Fahd University of Petroleum and Minerals (KFUPM), Dhahran, Saudi Arabia. His research interests include wireless sensor networks, real-time systems, cyber threat intelligence, and intrusion detection.
\end{IEEEbiography}

\vspace{-0.45pt}
\begin{IEEEbiography}[{\includegraphics[width=1in,height=1.25in,clip,keepaspectratio]{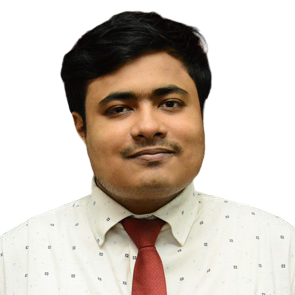}}]{Md Sakibul Islam} received his B.Sc. degree in Computer Science and Engineering from the Islamic University of Technology, Dhaka, Bangladesh, M.Sc. Degree from Erasmus Mundus GENIAL. He is currently pursuing his Ph.D. degree in Computer networks at King Fahd University of Petroleum and Minerals (KFUPM), Dhahran, Saudi Arabia.
\end{IEEEbiography}

\vspace{-0.45pt}
\begin{IEEEbiography}[{\includegraphics[width=1in,height=1.25in,clip,keepaspectratio]{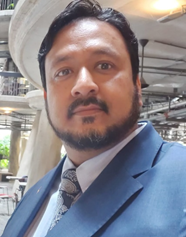}}]{Basem Al-madani} received the B.Sc. degree in computer engineering from the King Fahd University of Petroleum and Minerals (KFUPM), in 1997, and the M.Sc. degree in industrial automation and the Ph.D. degree from the Institute for Automation, Montan University of Leoben (MUL), Austria, in 1999 and 2005, respectively. He joined the SABIC International Team to manage the Industrial Automation Project in Vienna, Austria. He joined KNAPP Systems Integration in 2001, Leoben, Austria, as a Logistics Automation Specialist. He was the Chairperson of the Computer Engineering Department, KFUPM, from 2009 and 2014, where he is currently the Director of the Alfozan Academy, for Leaders Development in Non-Profit Sector Program. His research interests include real-time systems integration, distributed systems, middleware, IIoT, and IR 4.0.
\end{IEEEbiography}

\vspace{-0.45pt}
\begin{IEEEbiography}[{\includegraphics[width=1in,height=1.25in,clip,keepaspectratio]{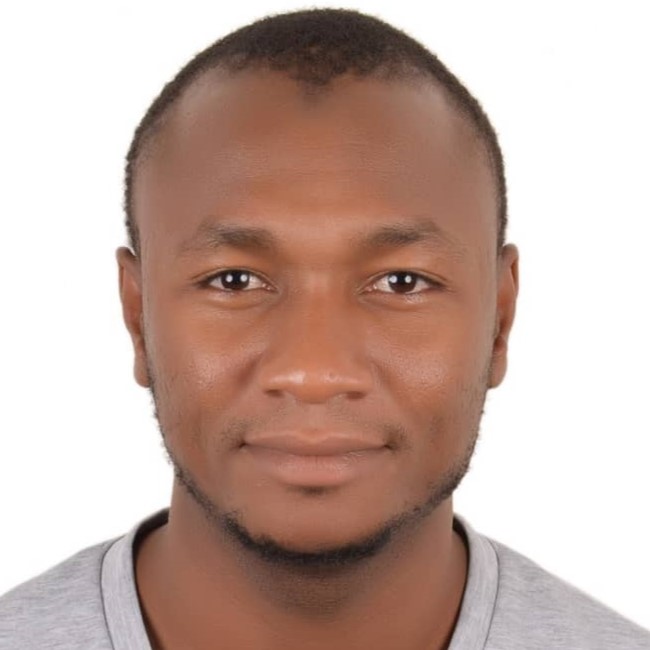}}]{Farouq Aliyu} (Senior Member, IEE) received a B.Eng. degree in computer engineering from Bayero University Kano (BUK), Kano, Nigeria, in 2010 and an M.S. degree in computer engineering from King Fahd University of Petroleum and Minerals (KFUPM), Dhahran, Saudi Arabia, in 2015. He also obtained his Ph.D. degree in computer engineering at KFUPM. He is currently a research engineer at the Applied Research Center (ARC) for Non-Profit and Social Development in KFUPM. His research interests include; Internet of Things (IoT) and Humanitarian Engineering.
\end{IEEEbiography}

\vspace{-0.45pt}
\begin{IEEEbiography}[{\includegraphics[width=1in,height=1.25in,clip,keepaspectratio]{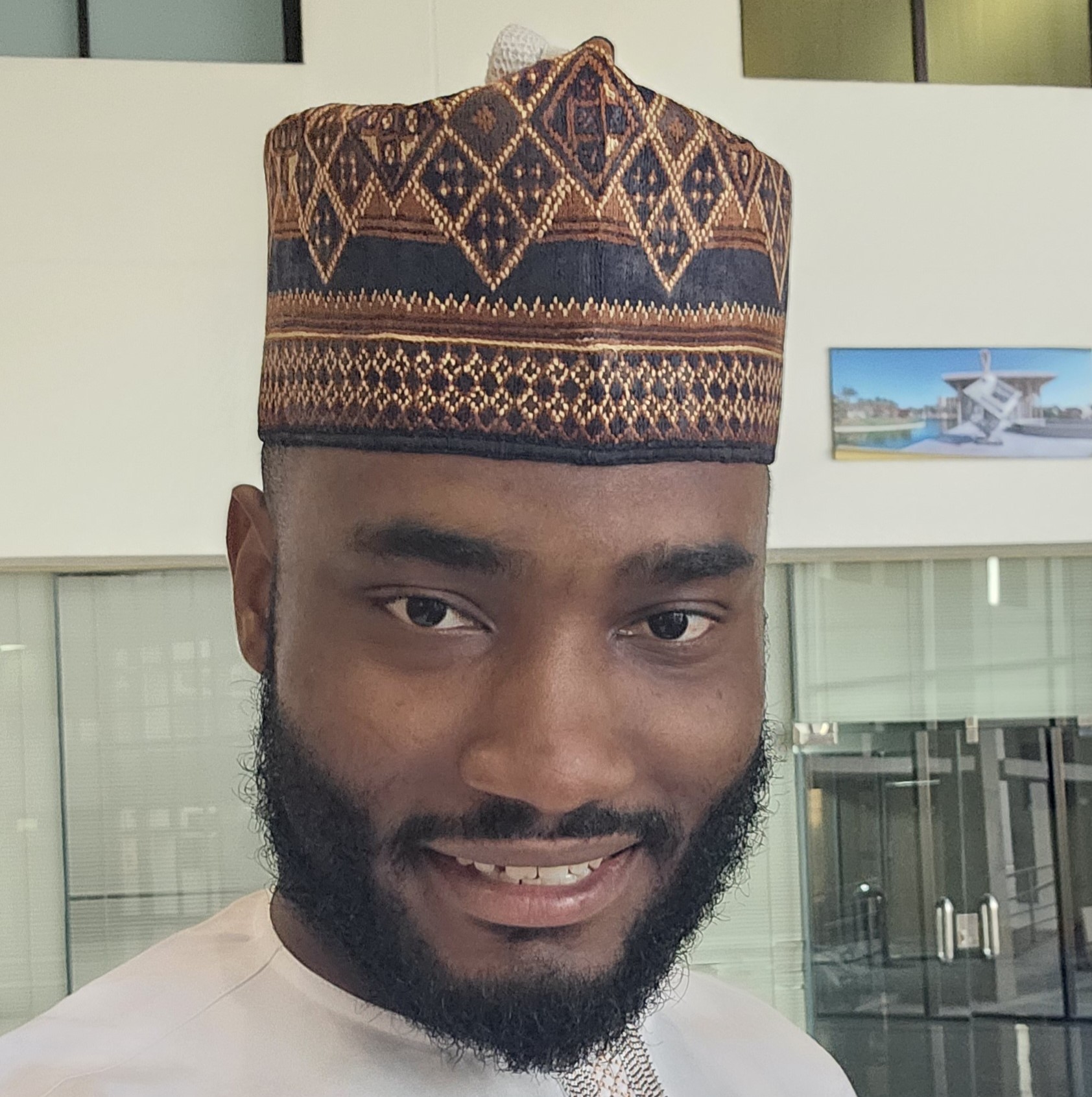}}]{Abdullahi Sani Shuaibu} (Student Member, IEEE) is a graduate student in computer engineering at King Fahd University of Petroleum and Minerals with a strong interest in interdisciplinary research at the intersection of artificial intelligence, robotics, and electrical and computer systems. He earned his bachelor's degree in computer engineering from the Federal University of Technology, Minna.
\end{IEEEbiography}

\vfill
\end{document}